\documentclass[%
 reprint,
groupedaddress,
 amsmath,amssymb,
 longbibliography,
 aps,
pra,
]{revtex4-1}

\usepackage{booktabs}
\usepackage{graphicx}% Include figure files
\usepackage{svg}
\usepackage{dsfont}
\usepackage{dcolumn}% Align table columns on decimal point
\usepackage{bm}% bold math
\usepackage{lipsum}
\usepackage{color}
\usepackage{xcolor}
\usepackage[normalem]{ulem}
\usepackage{simplewick}
\usepackage{qcircuit}
\usepackage{braket}
\usepackage{float}
\usepackage{multirow}
\usepackage{tikz}
\usepackage[utf8]{inputenc}

\usepackage{braket}
\usepackage{subfig}

\newcommand{\R}{\mathbb{R}} % real numbers
\newcommand{\C}{\mathbb{C}} % complex numbers
\newcommand{\hc}{\text{h.c.}} % complex numbers
\newcommand{\hpsi}{\hat{\psi}}
\newcommand{\hbpsi}{\hat{\bar{\psi}}}

\newcommand{\llink}{{(x, k)}}
\newcommand{\np}{n_\text{pauli}}
\newcommand{\tx}[1]{\text{#1}}

\renewcommand{\arraystretch}{1.2}
\newcommand{\ra}[1]{\renewcommand{\arraystretch}{#1}}

\newcommand{\rvline}{\hspace*{-\arraycolsep}\vline\hspace*{-\arraycolsep}}
\usepackage{makecell}
\usepackage{physics}

\begin{document}

\title{Toward scalable  simulations of  Lattice Gauge Theories   on  quantum computers}

\author{Simon V. Mathis}
\affiliation{IBM Research GmbH, Zurich Research Laboratory, S\"aumerstrasse 4, 8803 R\"uschlikon, Switzerland}

\author{Guglielmo Mazzola}
\affiliation{IBM Research GmbH, Zurich Research Laboratory, S\"aumerstrasse 4, 8803 R\"uschlikon, Switzerland}

\author{Ivano Tavernelli}
\affiliation{IBM Research GmbH, Zurich Research Laboratory, S\"aumerstrasse 4, 8803 R\"uschlikon, Switzerland}

\date{\today}
             
\begin{abstract}

    The simulation of real-time dynamics in lattice gauge theories is particularly hard for classical computing due to the exponential scaling of the required resources. On the other hand, quantum algorithms can potentially perform the same calculation with a polynomial dependence on the number of degrees of freedom.
    A precise estimation is however particularly challenging for the simulation of lattice gauge theories in arbitrary dimensions, where, gauge fields are dynamical variables, in addition to the particle fields.
    Moreover, there exist several choices for discretizing particles and gauge fields on a lattice, each of them coming at different prices in terms of qubit register size and circuit depth. 
    Here we provide a resource counting for real-time evolution of $U(1)$ gauge theories, such as Quantum Electrodynamics, on arbitrary dimension using the Wilson fermion representation for the particles, and the Quantum Link Model approach for the gauge fields.  We study the phenomena of flux-string breaking up to a genuine bi-dimensional model using classical simulations of the quantum circuits, and discuss the advantages of our discretization choice in simulation of more challenging $SU(N)$ gauge theories such as Quantum Chromodynamics.
    
\end{abstract}

\maketitle

\section{Introduction}

%Gauge theories represent an extremely useful theoretical tool in  physics, from particle physics to condensed matter.
Gauge theories are ubiquitous in physics, with applications ranging from fundamental particle theory to condensed matter.
The most popular gauge theories are the quantum field theories of the fundamental interactions, i.e. the Standard Model of particle physics~\cite{peskin1995introduction, schwartz2014quantum, weinberg1995quantum}, and quantum electrodynamics (QED)~\cite{dirac1927quantum,feynman2006qed}. 
Gauge theories play also a key role in statistical and condensed matter physics, e.g., in the study of high-temperature superconductivity, quantum spin liquids, topological quantum matter, and the fractional Hall effect to only name a few~\cite{affleck19882, kogut1979introduction,kleinert1989gauge,fradkin2013field, bruus2004many, PhysRevB.37.580, RevModPhys.89.025003, bernevig2013topological}. 
They find useful applications also in Quantum Information, e.g. the Kitaev's toric error correction code~\cite{Kitaev_2003, Gu_2014}.
Quantum Chromodynamics (QCD), a candidate theory to describe strong interactions, is perhaps the most studied and challenging gauge theory~\cite{halzen2008quark}.
While in the high energy limit perturbative approaches work well due to the so-called asymptotic freedom of the quarks~\cite{gross1973ultraviolet}, the low energy non-perturbative regime requires non-trivial numerical techniques. %due to confinement.

The most powerful method to understand nonperturbative QCD effects is the lattice gauge theory (LGT), introduced by Wilson~\cite{Wilson1974ConfinementQuarks} to describe the mechanism governing the confinement of quarks.
As the name suggests, LGT requires the discretization on a space-time lattice~\cite{kogut1983lattice,kogut1979introduction}.
However, the lattice QCD approach is almost prohibitively expensive and so far limited to the calculation of equilibrium properties, such as the equation of state at finite temperature, phase diagrams, quark masses, and scattering parameters~\cite{boyd1996thermodynamics, Tanabashi2018ReviewPhysics}.
Classical state of the art simulations of lattice QCD normally work in the path integral formulation on a Euclidean space-time lattice \cite{Smit2002IntroductionLattice,PhysRevLett.117.182001,PhysRevD.95.014507,PhysRevD.95.014507}. 
They are necessarily carried out on finite space-time boxes of spatial lengths $L_s = a N_s$ and temporal lengths of $L_t = a N_t$ with lattice spacing $a$ ($N_s$ sites in each spatial dimension and $N_t$ sites in the (imaginary) time direction). 
Typical volumes of $V_{lat} \sim [1$-$5]$~fm$^3$ are constrained from below by the allowable finite-size error and from above by the computational time taken to simulate the box with the finest lattice spacing. The largest lattices simulated to date feature $N_s = 144$ and $N_t = 288$ with total run-times of order of years~\cite{Tanabashi2018ReviewPhysics}.
As a result, the phase diagram of QCD remains largely unknown.
Moreover, the real time dynamics of proton collisions, flux string breaking or vacuum fluctuations cannot be 
 accessed using  present state-of-the art lattice QCD techniques based on the Lagrangian formalism, i.e. path-integral Monte Carlo methods~\cite{RevModPhys.82.1349}.

Real-time dynamics can in principle be formulated using a Hamiltonian formulation 
of lattice gauge theory~\cite{Kogut1975HamiltonianTheories} that is alternative to the Lagrangian approach. 
The Hamiltonian formulation retains only the spatial lattice dimensions and describes the time variable as a continuous parameter.
From a practical perspective, the primary restriction associated to this approach is represented by
its excessive computer memory requirements, as the resource needed to represent a general state on the discretized Hilbert space grows exponentially with the number of lattice sites.
While classically, compact representations of the quantum state based on tensor-networks have proven particularly successful in alleviating this issue\cite{banuls2013mass,buyens2014matrix,rico2014tensor,silvi2014lattice,kuhn2015non,pichler2016real} 
this shortcoming can be completely overcomed using quantum computers, where an exponentially increasing Hilbert space can be efficiently stored using linearly increasing resources, i.e. qubits~\cite{nielsen2002quantum}.
Moreover the  existence of a polynomial time complexity algorithm for simulating real time dynamics can offer a proven quantum advantage compared to equivalent classical computations~\cite{Feynman1982}, as pointed out in the seminal work in Ref.~\cite{jordan2012quantum} in the context of $\phi^4$ quantum field theory.

With the advent of quantum simulators, first experimental applications of quantum field theories on  quantum devices have appeared~\cite{buchler2005atomic,PhysRevLett.109.125302,Wiese2013UltracoldTheories, marcos2013superconducting,Mezzacapo2015Non-AbelianCircuits,PhysRevA.100.012320}. In 2016 the real-time dynamics of the Schwinger model~\cite{schwinger1962gauge}, a one-dimensional version of QED, were demonstrated for the first time on an ion-trap quantum computer~\cite{Monz2016Real-timeComputer, Muschik2017U1Simulators}.
%Shortly after, variational studies on the Schwinger model were conducted on ion-traps \cite{Kokail2018Self-VerifyingModel} and superconducting qubits \cite{KlcoQuantum-ClassicalComputers}.
Shortly after, several other studies have been put forward, either using different quantum computing architectures, such as superconducting qubits~\cite{KlcoQuantum-ClassicalComputers}, or following a time-independent variational approach~\cite{Kokail2018Self-VerifyingModel}.
While impressive, the scalability of these approaches is still unclear.
%are not scalable toward arbitrary dimensions. 
Indeed, they either rely on the specifics of one-dimensional gauge theories, which allow to integrate out all gauge field degrees of freedom entirely~\cite{Monz2016Real-timeComputer, Muschik2017U1Simulators, Kokail2018Self-VerifyingModel}, or require a classical preprocessing step of exponential time complexity~\cite{KlcoQuantum-ClassicalComputers,PhysRevA.100.012320}. 
On the other side, theoretical proposals for analog and digital quantum simulation of $SU(N)$ gauge theories in arbitrary dimensions do not detail the implementations down to a qubit level and an assessment of the required resources (gates, qubits)~\cite{byrnes2006simulating, PhysRevD.100.034518} is restricted to pure-gauge Hamiltonians~\cite{PhysRevLett.115.240502} or is tailored to cold-atom quantum simulators~\cite{Zohar2014FormulationSimulations,Bender2018DigitalDimensions,PhysRevResearch.2.013288}.
We refer to the recent Ref.~\cite{banuls2020review} for a review of  LGT quantum simulations using analog quantum computing devices, which are beyond the scope of this work, and to Ref.~\cite{dalmonte2016lattice} for a review on LGT and tensor network states.

Given the steady progress towards building larger digital quantum computers, it is highly desirable to identify the most efficient pathway to scale up quantum simulations of gauge theories, together with a precise estimation of the total resources needed.
In this work, we assess the scaling of the required qubit and gate resources to implement the $U(1)$ lattice gauge theory Hamiltonian for three different fermion-to-qubit mappings and two different gauge-field-to-qubit representations. This assessment is the first of its kind for a full digital quantum simulation of $U(1)$ lattice gauge theories with dynamical fermions, and it is relevant for both real-time evolution algorithms as well as variational approaches for ground state properties.
Moreover, we also provide directional guidance to the order of magnitude of the requirements for more complex gauge theories with non-Abelian Lie groups.

The paper is organized as follows. In Sect.~\ref{2:theoryclassical} we introduce QED in its Hamiltonian formulation and the most popular procedures to discretize the space coordinates on a lattice, while Sect.~\ref{3:qubits} deals with the mapping of the particle and field operators of the discretized QED Hamiltonian into the qubit space, using Wilson fermions and the quantum link model to map the gauge field.
In Sect.~\ref{sec:res_qed} we evaluate the scaling of the resource: number of qubits and the number of terms contributing to the qubit Hamiltonians, as a function of the size of the system and of the discretization parameters.
We then present in Sect.~\ref{5:real_time} an estimate of the circuit requirements in terms of number of gates and overall circuit depth, which are necessary to implement the real-time evolution algorithm.
Finally, in Sect.~\ref{6:example} we report some illustrative simulations of the dynamics of one-dimensional $U(1)$ model system,
%examples of typical problems that can be simulated with the proposed approach, 
while in Sect.~\ref{7:conclusion} we briefly discuss the applicability of the proposed real-time dynamics algorithm 
in near-future quantum computers.

\section{Hamiltonian formulation of U(1) lattice gauge theories}
\label{2:theoryclassical}

Quantum Electrodynamics (QED) is the relativistic quantum field theory that describes the interactions of charged particles with light.  QED is an instance of a $U(1)$ gauge theory, the simplest type of continuous gauge theory, and it is a cornerstone of the Standard Model.
$U(1)$ gauge theories are also applied in condensed matter physics, with applications ranging from the Hubbard model~\cite{PhysRevLett.95.036403}, spin liquids~\cite{RevModPhys.89.025003} and oxide superconductors~\cite{PhysRevB.37.580} to only name a few.

In this work, we consider lattice QED (LQED) as prototypical example of $U(1)$ lattice gauge theory.
While LQED, is one way to access the non-perturbative regime of QED,  nowadays, the main interest in LQED lies in its role as a test bed for simulation algorithms of more computationally involved gauge theories such as lattice QCD (characterized by the non-Abelian $SU(3)$ local gauge symmetry).
 
%{\bf Hamiltonian formulation.} 
Our analysis starts from the Hamiltonian formulation of continuous QED.
%The main reason to do so is not only to be instructive, but also to show how there are several possibilities available to discretize the theory on the lattice.
This is motivated by the need to develop a LGT in which only the space is discretized while the time variable remains continuous. 
Several discretization procedures are possible within this framework and different solutions, e.g. for the fermionic fields, have been adopted in state-of-the art implementations in \emph{classical} computers (which, for reference, are primarily based on the \emph{Ginsparg-Wilson domain wall fermion} approach~\cite{luscher1998exact}). 
Here, we will make use of this flexibility to devise the most suitable strategy for a scalable implementation of LGT suited for the new quantum computing paradigm.
 
We start by defining the operator valued fermionic Dirac spinor fields $\hat\psi(x)$ and $\hat{\bar{\psi}}= \hat\psi^\dagger(x) \gamma^0$, which in the context of QED represent the electron field. The number of spinor components equals $2^{d/2}$ for even $d$ and $2^{(d+1)/2}$ for odd $d$.
$\gamma^0$ is an element of the Dirac matrices $\{\gamma^{\mu}\}$ that form a $4 \times 4$ representation of the Clifford algebra $\text{Cl}(1,d)$~\cite{peskin2018introduction}.
The fermionic spinor fields obey equal-time canonical commutation relations:
\begin{align}
\label{4:psi_commutation}
\begin{gathered}
    \lbrace \hat{\psi}(x), \hat{\psi}(y) \rbrace = 0\\
    \lbrace \hat{\psi}^\dagger(x), \hat{\psi}^\dagger(y) \rbrace = 0  \\
    \lbrace \hat{\psi}(x), \hat{\psi}^\dagger(y) \rbrace = \delta^{(d)}(x-y).
\end{gathered}
\end{align}
The gauge field $\hat A_\mu(x)$ describes the photon field and the electromagnetic field tensor $\hat F_{\mu \nu}(x)$ is defined as 
\begin{equation}
    \label{4:eq:field_tensor}
   \hat F_{\mu \nu} (x) = \partial_\mu \hat A_\nu (x) - \partial_\nu \hat A_\mu(x).
\end{equation}
The physically measurable electromagnetic fields are defined from the electromagnetic field tensor as follows, $\hat E_k(x) = \hat F_{0k}(x)$ and $\hat B_i(x) = - (1/2) \varepsilon_{ijk} \hat F_{jk}(x)$.
The canonical commutation relations for the bosonic
gauge field operators $\hat{A_i}, \hat{E_j}$ are 
\begin{align}
    \label{4:eq:commutations}
    %\begin{gathered}
    \left[\hat{A}_i(x), \hat{A}_j(y) \right] &= 0 \\
    \left[\hat{E}_i(x), \hat{E}_j(y) \right] &= 0 \\
    \left[\hat{A}_i(x), \hat{E}_j(y) \right] &= -i\delta_{ij} \delta^{(d)}(x-y)~,
   % \end{gathered}.
\end{align}
where $\delta_{ij}$ is the Kronecker delta and $\delta(x)$ is the Dirac delta function.

The Hamiltonian operator of continuum QED reads~\cite{weinberg_1995}
%\begin{align}
\begin{multline}
    \label{4:eq:continuum_H_operator}
    \hat{H} = \int d x^d 
    \big( -\hat{\bar{\psi}} i\gamma^k \left[\partial_k - i q \hat{A}_k \right] \hat{\psi}+ \\
    + m \hat{\bar{\psi}} \hat{\psi} + \dfrac{1}{2} \hat{E}_i \hat{E}_i + \dfrac{1}{4} \hat{F}^{ij} \hat{F}_{ij} \big) ~~,
\end{multline}
%\end{align}
where the parameter $q$ is the electron charge ($q=-e,~e>0$) and $m$ is the electron mass. Summation of equal indices is assumed.

\subsection{Lattice discretization}
\begin{figure}[htb]
    \centering
    \includegraphics[scale=0.8]{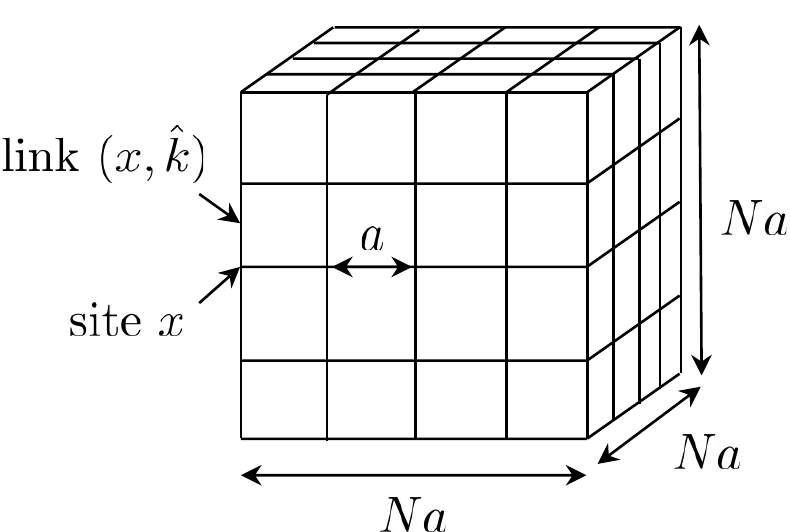}
    \caption{A hypercubic spatial lattice in $d$ dimensions with lattice spacing $a$ and $N$ lattice 
    sites (vertices) in each direction. The fermionic fields $\hat{\psi}_x$ will reside on the lattice sites and the gauge fields will occupy the links between the lattice sites as explained in the main text.}
    \label{4:fig:lattice}
\end{figure}

For the discretization of the continuum space, we follow the ideas of Wilson~\cite{Wilson1974ConfinementQuarks} and create
a lattice model of QED which implements the exact $U(1)$ symmetry at all lattice spacings. Lorentz invariance is however lost by
the lattice approach and is only recovered in the continuum limit.
To proceed, we slice the spatial region of interest into a (hyper)cubic lattice $\Gamma$ of lattice spacing $a$ with $N$ sites per lattice direction.
A meaningful setup features a lattice spacing $a$ roughly one order of magnitude smaller of the typical correlation length or less, while the lattice size should be larger than 3 times the correlation length or more.
An illustration is given in Fig.~\ref{4:fig:lattice}. 
The total volume of the lattice is $V=(Na)^d$. 
The fermionic fields are defined on the lattice \emph{sites}, while the gauge fields reside the \emph{links} connecting two neighboring lattice sites. 
An arbitrary site on the lattice is denoted by a real $d$-tuple $x = (x^1, \dots, x^d)= (n^1 a, \dots n^d a)$ where $x^k = n^k a$ for integers $n^k \in \lbrace 0, \dots N-1 \rbrace$. 
The unit lattice vectors in direction $k$ are denoted by $\hat{k}$. 
A link is denoted by the tuple $(x, \hat{k})$ (the same link is equivalently accessed by $(x+a\hat{k}, -\hat{k})$).

\subsection{Discretization of the fermionic fields}
The discretization of the QED Hamiltonian poses also important challenges.
%Discretizing the Hamiltonian on the spatial lattice is not as trivial as it may seem. 
A well know issue occurs already with the \emph{kinetic} term of Eq.~\ref{4:eq:continuum_H_operator}, i.e. the \emph{fermionic doubling problem}~\cite{Susskind1977LatticeFermions,rothe2005lattice,RevModPhys.84.449}, which consists in the appearance of spurious $2^d$ particles (called \emph{flavors}) for each physical one.
%In Appendix \ref{app:doubling} we provide an illustrative derivation of this issue.
The problem has been discussed in-depth in the Nielsen-Ninomiya No-Go theorem~\cite{Nielsen1981AbsenceTheory,itzykson1989statistical}, which, in short, states that in order to get rid of the fermion doublers one must sacrifice at least one of the following symmetries of the continuum Hamiltonian: hermiticity, locality, translational symmetry or the chiral symmetry for vanishing fermion mass $m$. 
Different workarounds for this issue can therefore be categorized according to which of the above symmetries they violate.
Here, we will focus on two well-known strategies, the so-called \emph{Wilson fermion} method, which breaks chiral symmetry for vanishing fermion mass, and the \emph{staggered fermion} method, which breaks translational symmetry instead. 

{\bf Wilson fermions.} 
The Wilson fermion method was contrived in the late 1970s~\cite{Wilson1974ConfinementQuarks} as a strategy to deal with fermion doubling. It introduces a momentum dependent mass term ($\partial_k$ here denotes the discretized derivative)
\begin{align}
   \hat{H}_\text{wilson} &= \dfrac{ar}{2} \sum\limits_{x,k} (\partial_k \hbpsi^\dagger(x)) (\partial_k \hpsi(x)) \notag \\
   &= -\dfrac{ar}{2} \sum\limits_{x,k} \hbpsi^\dagger(x) \partial^k \partial_k \hpsi \notag \\
   &= -\dfrac{ar}{2} \sum\limits_{x,k} \hbpsi^\dagger_x \left(\dfrac{\hpsi_{x+a\hat{k}} - 2 \hpsi_x + \hpsi_{x-a\hat{k}}}{a^2} \right) 
\end{align}
to the Hamiltonian, that vanishes linearly with $a$ in the continuum limit for sufficiently smooth fermionic fields $\hat{\psi}$. 
The dimensionless parameter $r$ regulates the strength of the Wilson correction and is typically set to $r=1$.

The excitations at the corner of the Brillouin zone (i.e. at wave-vector $k=\pi/a$) carry now the energy $2 r/a$
%\begin{equation}
 %   E_{+} = m + \dfrac{2 r}{a} \propto \dfrac{r}{a},
%\end{equation}
and is therefore removed from the low-lying energy spectrum in the continuum limit.

The Wilson fermion method displays several merits: 
\emph{i}) It is conceptually simple, allowing to straightforwardly transfer continuum observables to their counterparts on the lattice; \emph{ii}) It can treat an arbitrary amount of fermion flavors. 
On the downside, Wilson fermions break chiral symmetry for the vanishing mass limit~\cite{Smit2002IntroductionLattice, Zache2018QuantumFermions}. 
In lattice QCD applications, this makes it hard to study the regime of small quark masses (close to the mass-less limit) in numerical simulations or to simulate the spontaneous breakdown of chiral symmetry on the lattice~\cite{rothe2005lattice}. 
In addition, Wilson fermions converge to the continuum only with order $\mathcal{O}({a})$, 
%(cfn. Appendix \ref{app:doubling} )
%as seen in \cref{4:eq:wilson_fermions_around_zero}
although this can be improved by adding correction terms to the Hamiltonian, eliminating the leading order terms in $a$ at the expense of increasing the Hamiltonian's complexity~\cite{sheikholeslami1985improved,Tanabashi2018ReviewPhysics}.

\bigskip
{\bf Staggered fermions.} 
This approach distributes the fermionic degrees of freedom on different lattice sites, breaking the one-to-one correspondence between points in the lattice and points in the physical space.
Thereby the lattice spacing is increased and the Brillouin zone reduced~\cite{Kogut1975HamiltonianTheories,Susskind1977LatticeFermions, rothe2005lattice}.
Compared to Wilson fermions, the staggered fermions have the advantage of converging to the continuum limit faster, with order $\mathcal{O}({a^2})$.
While this method preserves chiral symmetry, it breaks the original translational symmetry by the lattice constant $a$. 
%As the fermionic components are distributed over several lattice sites, one lattice point does not represent a single space-time point anymore. 
%This makes staggered fermions conceptually challenging and their relation to the continuum fermions becomes harder to interpret. 
%This construction requires us to simulate more links between these lattice sites. 
Concerning its implementation for $2$-component spinors, full staggering requires $2^d$ times as many links as needed in the case of Wilson fermions. 
Finally, staggered fermions in $d$ dimensions only reduce the $2^d$ doublers perfectly when there are $2^d$ fermionic components to distribute on the lattice. 
This restricts the number of fermionic flavors that can be simulated in $d$ dimensions.
On the other hand, if only partial staggering is applied~\cite{Zache2018QuantumFermions, rothe2005lattice} a theoretically ill-founded rooting procedure must be used to recover the continuum result.

\bigskip
{\bf Quantum computing implementations.} 
As a matter of fact, all current LGT implementations on quantum computers focused exclusively on the staggered fermion approach in one dimension~\cite{Monz2016Real-timeComputer, Muschik2017U1Simulators, KlcoQuantum-ClassicalComputers}, where either all gauge fields on the links are integrated out~\cite{Monz2016Real-timeComputer, Muschik2017U1Simulators} or exponentially expensive pre-computations become necessary~\cite{KlcoQuantum-ClassicalComputers}. 
For dimensions higher than $d=1$ the full elimination of gauge fields is not possible and as a consequence the staggered fermions formulation will require the number of links to be simulated to increase by a factor $2^d$ in $d$ dimensions. 
As we will see in Sec.~\ref{3:qubits}, this translates into a significantly increase in the number of qubits and gate operations required in quantum simulations for a given lattice size.
Given that under these conditions the calculation of the links will become the computationally most expensive part (cf. Sect. \ref{3:qubits}), we conclude that the Wilson fermions approach will give the most promising implementation of fermionic LGT in near-future quantum computers.
The advantage of adopting the Wilson fermion representation has also recently been discussed in Ref.~\cite{Zache2018QuantumFermions} in the context of cold-atoms based quantum simulations.
 
\subsection{Discretization of the gauge fields}

The naive discretization of the kinetic term in Eq.~\ref{4:eq:continuum_H_operator} is clearly not gauge invariant.
To ensure invariance, the full covariant derivative term $D_k(x)=\partial_k - i q \hat{A}_k$ of the fermionic fields,   $\psi^\dagger(x) D_k(x) \psi(x)$, must be considered at once~\cite{Wilson1974ConfinementQuarks}. 
In fact, the covariant derivative generates strictly gauge invariant combinations and corresponds to an infinitesimal version of a parallel transport. 
%The finite difference version of the 
The parallel transport from a spacetime point $x$ to a nearest neighbouring point in direction $\hat{k}$ at distance $a$ is performed by the operator
\begin{equation}
    \label{4:eq:parallel_transport}
    \hat U(x, k) = \exp \left[-iq\int_{x_k}^{x_k+a} d y_k \hat A_k(y) \right] .
\end{equation}
With $\hat{U}(x, k) \in U(1)$ the combination $\hat \psi^\prime{}^\dagger(x) \hat U^\prime \llink \hat \psi^\prime(x+a\hat{k})$ is  
invariant under the gauge transformations.
For sake of readability we drop here and in the following the versor notation $\hat k$ in $\hat U(x, k)$. 
Wilson's key idea to adopt the finite parallel transporter $\hat U\llink$ instead of the infinitesimal parallel transporter $A_k(x)$  as LGT variable led to the Ansatz
\begin{equation}
    \label{4:eq:discrete_covariant_derivative}
    D_k(x)\hat \psi(x) \longrightarrow \dfrac{\hat U_\llink  \hat \psi_{x+a\hat{k}} -\hat U_{(x, -k)} \hat \psi_{x-a\hat{k}} }{2a}
\end{equation}
for the discretized covariant derivative.
This discretization reduces to the covariant derivative in the continuum limit and implements exact $U(1)$ gauge invariance, regardless of the lattice spacing $a$, provided that the $\hat A_k(x)$ are slowly varying fields over the length scale of a lattice spacing (see Appendix \ref{app:gauge} for the derivation).

\begin{figure}[htb] %this figure will be at the right
    \centering
    \includegraphics[scale=0.8]{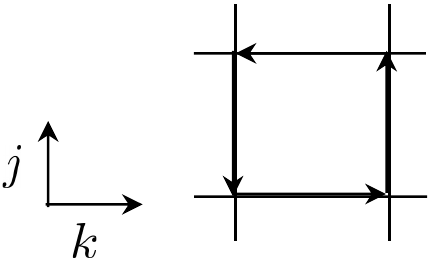}
    \caption{The plaquette term $\hat{U}_{(x,kj)}$ is a parallel transporter for a closed loop along edges in directions $k$ and $j$ moving from $x$.}
    \label{4:fig:plaquette}
\end{figure}

The next step consist in writing the magnetic field energy term of the Hamiltonian in Eq.~\ref{4:eq:continuum_H_operator} as a function of the parallel transporters $ \hat U$.
In the Appendix \ref{app:gauge} we show that this term can be written as
\begin{equation}
    \label{4:eq:magnetic_field_contribution}
    \dfrac{1}{4} \hat F_{kj}(x) \hat F^{kj}(x)  = \dfrac{1}{4 q^2 a^4} \sum\limits_{j<k} 2 - \left(\hat U_{(x, kj)} +\hat  U^\dagger_{(x, kj)} \right)
    + \mathcal{O}({a^2})~,
\end{equation}
where $\hat U_{(x, kj)}$ is defined as the parallel transporter for a closed loop along four edges (shown in Fig.~\ref{4:fig:plaquette} and usually called {\it{plaquette}}):
\begin{align}
\label{eq:plaq}
    \hat U_{(x, kj)} &= \hat U_{(x,k)} U_{(x+k,j)} \hat U_{(x+k+j,-k)} \hat U_{(x+j,-j)} \notag \\
    &=\hat U_{(x,k)}\hat U_{(x+k,j)}\hat U^\dagger_{(x+j,k)}\hat U^\dagger_{(x,j)},
\end{align}

The canonical commutation relations for the operators $\hat{U}_\llink$ are obtained from the commutation relations for $\hat{A}_k(x)$ and $\hat{E}_k(x)$, and read
\begin{align}
    %\begin{gathered}
    \left[\hat{U}_{(x, i)}, \hat{U}_{(y, j)} \right] &= 0  \label{4:eq:commutations_UE_1} \, ,\\
    \left[\hat{E}_{(x, i)}, \hat{E}_{(y, j)} \right] &= 0 \label{4:eq:commutations_UE_2} \, , \\
    \left[\hat{E}_{(x, i)}, \hat{U}_{(y, j)} \right] &= e \delta_{ij} \delta_{xy} \hat{U}_{(x, i)} \, . \label{4:eq:commutations_UE_3}
   % \end{gathered}.
\end{align}
That is, operators have the non-vanishing commutator $[\hat{E}, \hat{U}]=e \hat{U}$ only if referred to the same link.
The physical interpretation of this commutation relation is apparent if we look at an eigenstate $\ket{E}$ of $\hat{E}$ with eigenvalue $E$ on a single link. 
Then the relation in Eq.\ref{4:eq:commutations_UE_3} implies (suppressing the subscript $(x, \hat{k})$ for the time being)
\begin{equation}
    \hat{E}(\hat{U}\ket{E}) = (E + e) \hat{U} \ket{E} \, 
\end{equation}
meaning that the state $\hat{U}\ket{E}$ is again an eigenstate of $\hat{E}$ with flux eigenvalue $E + e$. 
Therefore, the operator $\hat{U}$ ($\hat{U}^\dagger$) acts as electric flux raising (lowering) operator for the eigenstates of $\hat{E}$ by one unit $e$ of electric flux. To take into account a constant background electric field, we write $\hat{E}_{(x, i)} \to \hat{E}_{(x, i)} + \theta_i$ where $\theta_i$ is a constant electric field along dimension $i$.

Combining all of the terms and dropping contributions which only lead to a shift in the total energy, the Wilson corrected lattice Hamiltonian becomes
\begin{widetext}
\begin{equation}
\begin{aligned}
    \label{4:eq:H_cqed_rescaled}
    \hat{H} &= 
    \hat{H}_\text{hopp} + \hat{H}_\text{mass} + \hat{H}_\text{wilson} + \hat{H}_\text{elec} + \hat{H}_\text{plaq}\\
    & =a^d \Biggl[ \sum\limits_\text{sites} \sum\limits_k \dfrac{1}{2a} \left(\hbpsi_x [i \gamma^k + r]
    \hat{U}_\llink \hpsi_{x+\hat{k}} + \hc \right)
    + \sum\limits_\text{sites} \left(m + \dfrac{r d}{a} \right) \hbpsi_x \hpsi_x \\
    &\hspace{1cm} + \dfrac{e^2}{2} \sum\limits_\text{links} \left(\hat{E}_\llink + \theta_k \right)^2  
    - \dfrac{1}{4 e^2} \sum\limits_\text{plaq.} (\hat{U}_\Box + \hat{U}^\dagger_\Box) \Biggr] \, .
\end{aligned}
\end{equation}
\end{widetext}
Note that we rescaled the fields as 
$\hat \psi_x \to \sqrt{a} \hat \psi$, $\hat E_\llink \to a \hat  E_\llink/e $ 
and 
$\hat U_\llink \to \hat  U_\llink/a$, 
and defined the short-hand notation $\hat{U}_\Box$ for the generic \emph{plaquette} operator of Eq.~\ref{eq:plaq}.

The first term of this Hamiltonian, $\hat{H}_\text{hopp}$, describes the hopping of fermionic excitations from one lattice site to 
another. The second term, $\hat{H}_\text{mass}$, gives fermionic excitations their mass. 
We shall use the name $\hat{H}_\text{elec}$ for the third and $\hat{H}_\text{plaq}$ for the fourth term and they describe the electric field and magnetic field energies respectively. 

During its evolution, a state $|\phi \rangle$ also need to  satisfy the Gauss law at any given time. 
The discretized version of the Gauss law is given by
\begin{equation}
    \hat{G}_x \ket{\phi} = \left[
    \sum\limits_{k=1}^d \left( \hat{E}_{x-\hat{k}, k} - \hat{E}_{x,k}\right) - q \hpsi_x^\dagger \hpsi_x 
    \right] \ket{\phi} = 0 \, .
    \label{4:eq:lattice_gauss_law}
\end{equation}
for each lattice site $x$. 
This constraint singles out a subsector of the Hilbertspace which we will call the ``physical Hilbert space'' $\mathcal{H}_\text{phys}$, and is difficult to implement exactly on a quantum computer. 
We propose to enforce this condition by adding the regulator term 
\begin{equation}
    \hat{H}_\text{Gauss} = \sum\limits_{x\in\Gamma} \hat{G}_x^2 \, .
    \label{4:eq:gauss_regulator}
\end{equation} 
In the effective Hamiltonian $\hat{H}_\text{eff}=\hat{H} + \lambda \hat{H}_\text{Gauss}$ the unphysical states receive an energy penalty that scales with the regularization parameter $\lambda$. 
For large enough values of $\lambda$ the low-energy spectrum of the effective Hamiltonian then converges to the low energy spectrum of $\hat{H}$ in the physical Hilbert space~\cite{dalmonte2016lattice}.

In future simulations, the effective gauge invariance principle will need to be complemented by error correction. The error correction methods will have to mitigate the effect of errors which kick the quantum computation outside the physical Hilbert space despite effective gauge invariance. 
Such error correction could be achieved for example with the help of ancilla qubits to periodically measure whether the Gauss law constraint at any given lattice vertex is satisfied~\cite{PhysRevA.99.042301}.

\section{Mapping the Hamiltonian to a qubit operator
\label{3:qubits}}

In order to use fermionic and gauge operators on a quantum computer, we need to map them to qubit operators, which, in turn, are expressed  in terms of Pauli strings. \\

The set of $n$-qubit Pauli strings $\mathcal{P}_n = \lbrace p_1 \otimes
p_2 \otimes \dots \otimes p_n~|~p_i \in \lbrace I, X, Y, Z \rbrace \rbrace$ consists of $|\mathcal{P}_n| = 4^n$ tensor products of $n$ Pauli operators
\begin{align}
      \label{eq:pauli_spin_operators}
    I \equiv \mathbb{I} = \left( \begin{array}{cc}
        1 & 0 \\
        0 & 1
    \end{array}
    \right),~~
    X \equiv \hat{\sigma}^x = \left( \begin{array}{cc}
        0 & 1 \\
        1 & 0
    \end{array}
    \right),\\
    Y \equiv \hat{\sigma}^y = \left( \begin{array}{cc}
        0 & -i \\
        i & 0
    \end{array}
    \right),~~
    Z \equiv \hat{\sigma}^z = \left( \begin{array}{cc}
        0 & 1 \\
        1 & 0
    \end{array}
    \right).  
\end{align}

%\end{equation*}
%For readability we omit the tensor product in the writing of Pauli strings, for example we write $XIZ$ for the Paulistring $X \otimes I \otimes Z$. The single qubit Pauli strings $\mathcal{P}_1$ form a basis of the $\R$-vectorspace of $2 \cross 2$ hermitian matrices and of the four dimensional $\C$-vectorspace of general $2 \cross 2$ matrices. This means that we can represent any hermitian (general) operator on a single qubit as a linear combination of these four Pauli operators with coefficients in $\R$ ($\C$). As is well established in linear algebra, we can then express any operator on an $n$-qubit Hilbertspace as a real (complex) linear combination of Pauli strings $P \in \mathcal{P}_n$. More
Formally, any operator $\hat{O}_n$ on an $n$-qubit Hilbert space $(\C^2)^{\otimes n}$ can be mapped to a matrix form as 
\begin{equation}
    \label{4:eq:paulis_decomposition}
    O_n = \sum\limits_{P \in \mathcal{P}_n} \lambda_P P\, , \hspace{1cm} \lambda_P \in \C \, .
\end{equation}
The coefficients of this sum can be obtained by taking the scalar product 
\begin{equation}
    \label{4:eq:paulis_scalar_product}
    \langle p_1 p_2 \dots p_n,  q_1 q_2 \dots q_n \rangle 
 = \dfrac{1}{2^{n/2}}
 \prod\limits_{k=1}^{n} \sqrt{\mathrm{Tr}(p_k^\dagger q_k)} \, .
\end{equation}
where $\{q_i\}$ are the elements of the tensor products that define the basis in $\mathcal{P}_n$.

\subsection{Qubit representation gauge field operators}

Since gauge theory groups are continuous, the corresponding single link Hilbert space $\mathcal{H}_l = L^2(U(1))$ is infinite dimensional and has to be truncated in numerical implementations.
A first truncation approach consists in finding a finite group that share similar properties as the continuous one. 
For $U(1)$ gauge theories, the candidate finite group is the cyclic group
$\mathcal{C}_n \cong \mathbb{Z}_n$ with $n$ elements~\cite{creutz1983monte, PhysRevA.90.042305, Notarnicola2015DiscreteQED, Zohar2016DigitalMatter, PhysRevLett.123.090501, ercolessi2018phase}  
However, this approach still lack of a 
rigorous proof on the convergence of $\mathcal{C}_n$ to $U(1)$ when increasing $n$. 
On the other hand, Abelian gauge theories with any finite discrete group $\mathbb{Z}_n$ do not fall in the same universality class as the $U(1)$ model and therefore do not approach the $U(1)$ gauge theory as the space-time continuum limit is taken.
A second truncation approach is the so called \emph{quantum link} formalism~\cite{Wiese2013UltracoldTheories}, which is the method of choice in this work (see Sect.~\ref{ss:qlm}).
Interestingly, we notice that the discretization issue is virtually not present in modern classical simulations as 64-bit double-precision floating point numbers guarantees sufficient numerical precision.
A recent series of papers investigates the discretization errors introduced in digitizing elements of the gauge group to a finite set, with particular focus on $SU(2)$ gauge theories, for which an extensive dataset can be extracted with Monte Carlo methods ~\cite{PhysRevA.99.062341,PhysRevD.100.114501}.
Yet, these studies do not provide a constructive method to encode the finite mesh at a qubit level.

\subsubsection{Quantum link model}
\label{ss:qlm}
The idea behind the quantum link model (QLM) approach is to find a finite dimensional gauge field Hilbert space and corresponding `discretized' operators for $\hat{E}$ and $\hat{U}$, such that the fundamental commutation relation $[\hat{E},\hat{U}]=e\hat{U}$ is fulfilled while $\hat{E}$ remains hermitian.
The convergence of the truncated QLM to the original single link Hilbert space $\mathcal{H}_l$ has been proven in Ref.~\cite{Chandrasekharan_1997}. 
Moreover, a generalization of QLM to the gauge groups $SU(N)$ and $U(N)$~\cite{Brower1999QCDModel} also exists.
For the case of a $U(1)$ gauge group, a QLM was constructed explicitly in Ref.~\cite{Chandrasekharan_1997} by replacing each link with a spin $S$ system. 
To further simplify the notation, in the following we will drop the link subscript $\llink$, keeping in mind that operators at different links commute. 
On each single link, we map
\begin{equation}
\begin{aligned}
    \label{4:eq:qlm_mapping}
    \hat{E} &\to e\hat{S}^z \\
    \hat{U} &\to [S(S+1)]^{-1/2} \hat{S}^+.
\end{aligned} 
\end{equation}
where $\hat{S}^z$ and $\hat{S}^+$ are the spin $z$ and the spin raising operator respectively and $S(S+1)$ is the eigenvalue of the $\hat{S}^2$ operator. 
The commutation relation $[\hat{E}, \hat{U}] = e\hat{U}$ is then retained as a quick calculation with the spin commutation relations
\begin{equation}
    \label{4:eq:spin_commutation relations}
    [\hat{S}^\alpha, \hat{S}^\beta] = i \varepsilon_{\alpha \beta \gamma} \hat{S}^\gamma
\end{equation}
shows. 
%%%%%%%%%%%%%%%%%%%%%%
Here we use the definitions $\alpha, \beta, \gamma \in \lbrace x, y, z \rbrace$ and $\hat{S}^\pm = \hat{S}^x \pm i \hat{S}^y$. 
The map in Eq.~\ref{4:eq:qlm_mapping} replaces each link Hilbert space $\mathcal{H}_l$ with the one of a spin $S$ system with dimension $d_S=2S+1$. 
Since the relation $[\hat{E}, \hat{U}] = e \hat{U}$ also holds for the spin system $S$, the truncated lattice model preserves gauge invariant by construction, at the cost of scarifying the unitarity of the link operators, which now satisfy
\begin{equation}
    \label{4:eq:UU_commutator}
    [\hat{U}, \hat{U}^\dagger] = \dfrac{2}{e S(S+1)} \hat{E},
\end{equation}
instead of $[\hat{U}, \hat{U}^\dagger] = 0$.
However, note that the right-hand-side of Eq.~\ref{4:eq:UU_commutator} approaches zero as $S$ tends to infinity, restoring the desired property. 

\begin{figure}[htb]
    \centering
    \includegraphics[scale=0.6]{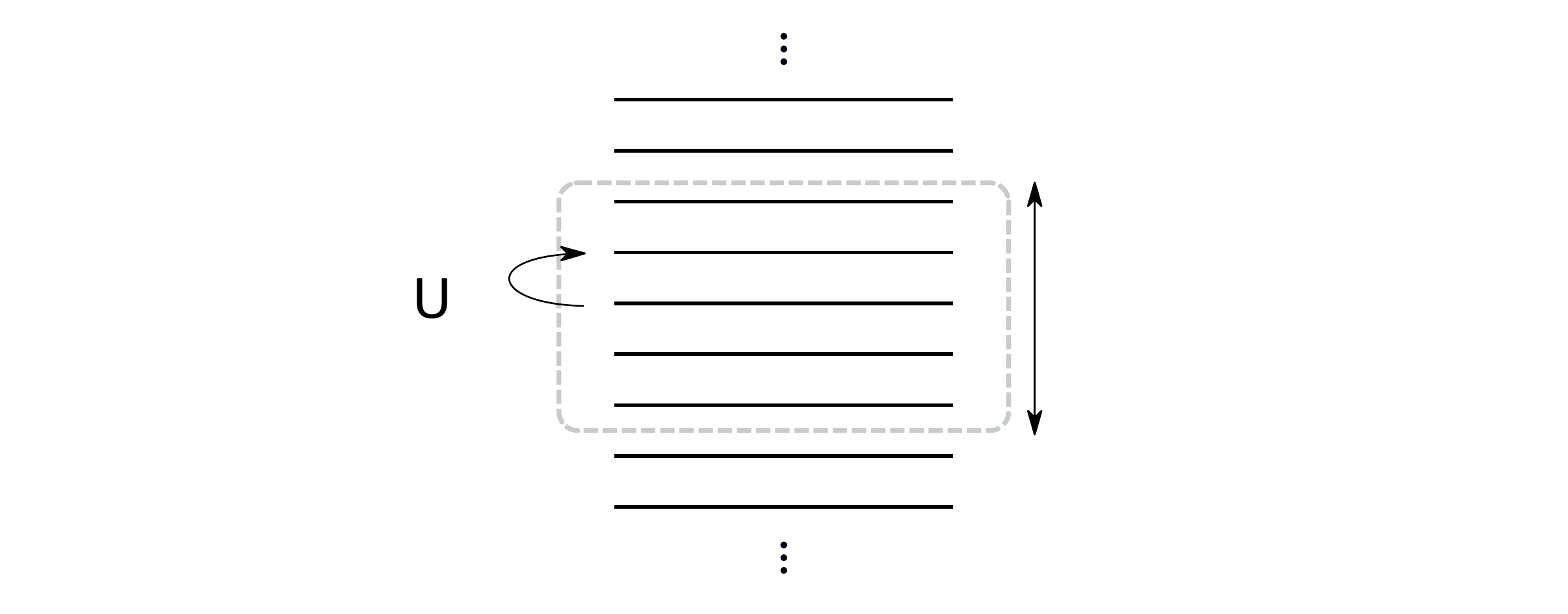}
    \caption{An illustration of the infinite flux ladder of eigenstates to the operator $\hat{E}$ and the window of size $d_S = 2S+1$ that selected by the quantum link model truncation.}
    \label{4:fig:flux_ladder}
\end{figure}

In the quantum link model, the flux basis is given by the $d_S=2S+1$ spin states
\begin{equation}
    \ket{-S}, \ket{-S+1}, \dots \ket{S-1}, \ket{S},
\end{equation}
which cut out a window with $d_S$ states (pictorially shown in Fig.~\ref{4:fig:flux_ladder}) from the infinite flux ladder of eigenstates of the $\hat{E}$ operator.

We can interpret the physical effects of the spin truncation value $S$ by means of the Gauss law in Eq.~\ref{4:eq:lattice_gauss_law}. The QLM cuts out a window of $d_S$ states from the infinite flux ladder and it allows therefore a flux increase of maximally $d_S$ units along any electric field line. 
If the given link is traversed by an uniform ($k$ independent) background field of intensity $\theta$ (Eq.~\ref{4:eq:H_cqed_rescaled}), this means that a QLM with $d_S > n_s/2 = n^d/2$ contains all possible states build around the offset flux value $\theta$, which satisfy the Gauss law. 
The spin truncation therefore does not introduce any further truncation error beyond the one associated to the lattice discretization. Modern high-precision calculations of lattice gauge theories on classical computers achieve convergence to the continuum limit with $n \sim 100$, which would correspond to truncation values of $S \sim \order{10^6}$ to reach the same accuracy. %completely eliminate any truncation error. 
However, states with large flux values (and consequently also a large number of particle pairs) incur a significant energy penalty from the field and mass terms in the Hamiltonian. 
As a consequence, %many of the
states which require a large truncation value $S$ %to be represented 
play a negligible role in the low-energy sector in which we are interested. 
In practice, even in the more complicated case of the $SU(3)$ gauge theory a mesh of $\order{10^4}$ elements for each $SU(3)$ gauge link was shown to be sufficient to reach the desired accuracy~\cite{PhysRevA.99.062341}.  
These results suggest that truncation values on the order of $S \sim \order{10^3} - \order{10^4}$ or less are likely large enough for the simpler case of $U(1)$ lattice gauge theories to eliminate truncation errors. %to neglect resulting truncation errors for precision calculations. 
%This hypothesis could be tested by an analogous calculation as in Ref.~\cite{PhysRevA.99.062341} for a $U(1)$ theory. 
%Note also that a QLM with $d_S > n/2$ includes states with all possible charge densities $\rho \leq q n^d / (2V)$ from the Hilbert space. 

\subsubsection{Spin $S$ to qubit mapping}
\label{4:ssec:spin_registers}
 
%For a general lattice gauge theory with an underlying $U(N)$ or $SU(N)$ symmetry, the relevant quantum link model would correspond to representations of $SU(2N)$~\cite{Brower1999QCDModel}.
%In the particular case of a $U(1)$ gauge theory, the QLM corresponds to a spin $S$ system \cite{Chandrasekharan_1997} (cf. Sec.~\ref{ss:qlm}).

Spin $S$ systems correspond to the irreducible representations of $SU(2)$ of dimension $d_S = 2S+1$, where $S$ is a positive half integer $S \in \lbrace 1/2, 1, 3/2, 2, \dots \rbrace$. 
The key operators on a spin $S$ system are a generalization of the Pauli $X$, $Y$, and $Z$ operators of the spin $1/2$ system. 
These operators are given by the representation matrices of the Lie algebra generators~\footnote{To be precise, the generators of $\mathfrak{su}(2)$ are anti-hermitian, so the hermitian spin operators correspond to $i$ times the representation matrix of the generators.} in the irreducible $\mathfrak{su}(2)$ representation to spin $S$. 
A computation via the highest-weight method \cite{Felder2016MathematischeVorlesungsskript} for $SU(2)$ yields
\begin{multline}
  \left(\hat{S}_x\right)_{lk} = \sqrt{S(S+1) - (S-l)(S-l+1)} \times \\
\times (\delta_{l,k+1} + \delta_{l, k-1})/2, \\
\left(\hat{S}_y\right)_{lk} = \sqrt{S(S+1) - (S-l)(S-l+1)} \times \\ 
\times (i\delta_{l,k+1} - i\delta_{l, k-1})/2, \\
\hat{S}_z = \text{diag}(S, S-1, \dots, -S+1, -S).  
\end{multline}
The spin raising and lowering operators $S_\pm$ are defined as $\hat{S}_\pm = (\hat{S}_x \pm i \hat{S}_y)/2$.

In the following, we consider here  two rather straightforward spin $S$ to qubit mappings, which we call the logarithmic and the linear encoding.

\bigskip
{\bf Logarithmic encoding.}
We define the logarithmic encoding of spin $S$ systems in qubits via the following
mapping of the $d_S$ eigenstates $\ket{m_s}, ~ m_s \in \lbrace -S, -S+1, \dots, S-1, S
\rbrace $ of the spin $z$ operator $\hat{S}_z$ to $n= \lceil \log_2{d_S} \rceil$ qubits:
\begin{equation}
    \begin{aligned}
    \label{4:eq:log_embedding}
    \ket{m_s = S} &\mapsto \ket{0}\\
    \ket{m_s = S-1} &\mapsto \ket{1}\\
    &\vdots \\
    \ket{m_s = -S+1} &\mapsto \ket{d_S-2}\\
    \ket{m_s = -S} &\mapsto \ket{d_S-1}
    \end{aligned}
\end{equation}
The Pauli string representation of the relevant spin operators is obtained by first embedding the spin operator into a matrix of size $2^n$ 
\begin{equation}
\label{4:eq:spin_in_matrix}
\hat{S}_{x,y,z}^\text{emb.} = 
\begin{pmatrix}
  \begin{matrix}
  \hat{S}_{x,y,z}
  \end{matrix}
  & \rvline & 0 \\
\hline
  0 & \rvline &
  \begin{matrix}
  \mathbb{I}
  \end{matrix}
\end{pmatrix}
\end{equation}
and then taking the Pauli scalar product (Eq.~\ref{4:eq:paulis_scalar_product}) of the embedded spin operators with the Pauli strings in $\mathcal{P}_n$ (cfn. Appendix~\ref{app:mixed} for results in the $S=1$ case).

\begin{figure}[htb]
    \centering
    \includegraphics[width=\columnwidth]{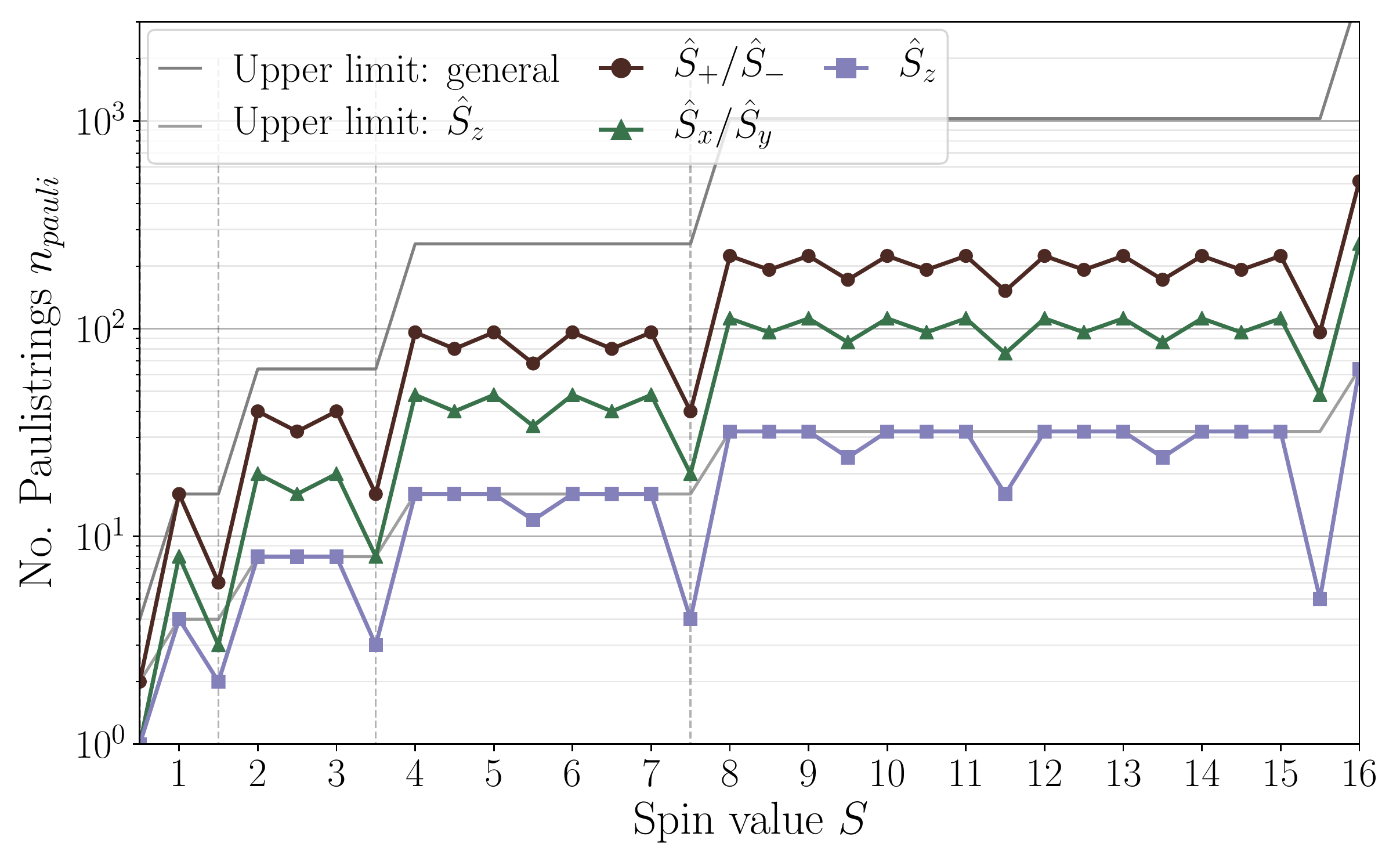}
    \caption{Number of Pauli strings required to represent the spin operators $\hat{S}_\pm, \hat{S}_x, \hat{S}_y, \hat{S}_z$ as qubit operators using the logarithmic encoding as a function of the total spin $S$. 
    The dashed vertical lines are a guide to the eyes for values of $S$ for which the spin system dimension $d_S = 2S+1$ is a power of two. (i.e., the so-called {\it perfectly representable} systems).}
    \label{4:fig:log_paulistring_scaling}
\end{figure}

We now briefly discuss the upper limits of the resource scaling for the logarithmic embedding as a function of $S$ and, in particular, the number of Pauli strings needed to implement a spin operator 
and the maximal support of these Pauli strings. The support being defined as the maximum number of non-trivial single-qubit $\{X,Y,Z\}$ Pauli operators in $\mathcal{P}_n$. 
The spin $z$ operator $\hat{S}_z$ is diagonal in the computational basis. %our encoding. 
As a consequence, it can only be made up by Pauli strings that exclusively contain the diagonal operators $I$ and $Z$. 
From this consideration, we obtain the straightforward upper bound 
\begin{equation}
    \label{4:eq:nz_upper_bound}
    n_\text{pauli}[\hat{S}_z] \leq 2^{\lceil \log_2{d_S} \rceil} \leq 2 d_S \, ,
\end{equation}
which is linear in the spin system's dimensionality $d_S$. 
For the tridiagonal spin $x$ and $y$ operators no similar restrictions apply. 
They can be composed by Pauli strings consisting of all four basic operators $I, X, Y, Z$. The corresponding loose upper bound
\begin{equation}
    \label{4:eq:nx_upper_bound}
    n_\text{pauli}[\hat{S}_{x,y}] \leq 4^{\lceil \log_2{d_S} \rceil} \leq 4 d_S^2
\end{equation}
is quadratic in the system's dimension. 
Note also that $n_\text{pauli}[\hat{S}_x] = n_\text{pauli}[\hat{S}_y]$, since $\hat{S}_x$ and $\hat{S}_y$ have the same off-diagonal structure. 
To test the looseness of these bounds, we studied, numerically, the exact number of required Pauli strings under the logarithmic mapping for small values of $S$. The result is displayed in Fig.~\ref{4:fig:log_paulistring_scaling}. 

Notably, the mapping of spin $S$ systems %where the dimensionality of the system is 
with dimension corresponding to a power of $2$ requires far less Pauli strings than predicted by the given upper bounds. 
We call such spin systems {\it{perfectly representable}}. 
They map perfectly to the computational basis states of $\log_2{d_S}$ qubits, leaving no unphysical states in the spectrum. 
To estimate the required scaling behavior, we perform a fit in the range of values of $d_S \leq 2^{10}$. 
A physically motivated fit function is given by $f(x)=a x^2 + b x^{\log_2{3}} + c x + d$, as each term of this function corresponds to a number of Pauli strings with support four, three, two and one, respectively. Such a fit yields
%\begin{multline}
 %   \label{eq:pauli_Sxy}
  %  n_\text{pauli}^{\text{log-perf.}}(\hat{S}_{x,y}) \sim 1.14 \cdot 10^{-20} d_S^2 + \\ + 1.68 \cdot 10^{-2} d_S^{\log_2{3}} + 1.85 d_S + 1.03.
%\end{multline}
\begin{equation}
        \label{eq:pauli_Sxy}
    n_\text{pauli}^{\text{log-perf.}}(\hat{S}_{x,y}) \sim    1.68 \cdot 10^{-2} d_S^{\log_2{3}} + 1.85 d_S + 1.03 \, 
\end{equation}
with vanishing quadratic coefficient.
The fit indicates a cross-over from the linear to the $log_2{3}\sim1.58$ regimes at $d_S \sim 3000$. The latter represents the asymptotic scaling as the quadratic regimes is always irrelevant for any reasonable value of $d_s$.
 We also note that Eq.~\ref{eq:pauli_Sxy} is an upper bound.
 %, as seen by fitting only part of the available data.

On the other hand, the mapping of perfectly representable $\hat{S}_z$ operators requires
\begin{equation}
    n_\text{pauli}^{\text{log-perf.}}(\hat{S}_{z}) = \log_2{d_S}
\end{equation}
Pauli strings with support 1 each (e.g. $ZII...I$, $IZI...I$). 

\bigskip
{\bf Linear encoding.} 
In the linear qubit encoding of spin $S$ systems, we map the eigenstates of the spin $\hat{S}_z$ operator to $n=d_S$ qubits:
\begin{equation}
    \begin{aligned}
    \label{4:eq:lin_embedding}
    \ket{m_s = S} &\mapsto \ket{00\dots001}\\
    \ket{m_s = S-1} &\mapsto \ket{00\dots010}\\
    &\vdots \\
    \ket{m_s = -S+1} &\mapsto \ket{01\dots000}\\
    \ket{m_s = -S} &\mapsto \underbrace{\ket{10\dots000}}_{d_S \text{ digits}}\\
    \end{aligned}
\end{equation}
To derive the form of the spin operators in this encoding, we first look at the spin raising operator $\hat{S}_+$. 
To generate the spin state $\ket{m_s+1}$ from $\ket{m_s}$ in the qubit encoding of Eq.~\ref{4:eq:lin_embedding} we need to perform the operation
\begin{multline}
      \label{4:eq:msplusminus}
    \sigma^+_{m_s+1}\sigma^-_{m_s} = X_{m_s}X_{m_s+1} + Y_{m_s}Y_{m_s+1}+ \\ + i(Y_{m_s}X_{m_s+1}- X_{m_s}Y_{m_s+1}) \,.  
\end{multline}
Here the operators $\sigma^\pm_{m_s}$ refer, respectively, to the lowering and raising operators on the qubit marking the $m_s$-th spin state, with
\begin{equation}
    \sigma^+ = \left(
    \begin{array}{cc}
        0 & 1 \\
        0 & 0
    \end{array}
    \right), \hspace{1cm} \sigma^- = \left(
    \begin{array}{cc}
        0 & 0 \\
        1 & 0
    \end{array}
    \right).
\end{equation}
From Eq.~\ref{4:eq:msplusminus} we can construct the general spin raising operator by adding these terms in a linear combination with the correct coefficients, which can be obtained via the highest-weight method~\cite{Felder2016MathematischeVorlesungsskript}. 
The encoded $\hat{S}_+$ operator is then
\begin{equation}
    \label{4:eq:lin_embedded_splus}
    \varepsilon(\hat{S}_+) = \sum\limits_{m_s = -S}^{S-1} \sqrt{S(S+1)-m_s(m_s+1)} \sigma^+_{m_s+1} \sigma^-_{m_s}.
\end{equation}
Consequently, the encoded spin $x,y,z$ operators are  
\begin{equation}
    \label{4:eq:lin_spin_xyz}
    \begin{aligned}
    \varepsilon(\hat{S}_x) = (\varepsilon(\hat{S}_+) + \varepsilon(\hat{S}_+)^\dagger)/2 \\
    \varepsilon(\hat{S}_y) = i(\varepsilon(\hat{S}_+) - \varepsilon(\hat{S}_+)^\dagger)/2 \\
    \varepsilon(\hat{S}_z) = \sum\limits_{m_s = -S}^S m_s (\sigma^z_{m_s} + \mathbb{I})
    \end{aligned}
\end{equation}

\begin{table*}\centering
\ra{1.2}

\begin{tabular}{@{}cccccccc@{}}\toprule
\hline %put by GM
\makecell{Spin S} & \multicolumn{3}{c}{Linear encoding} & \phantom{a}& \multicolumn{3}{c}{Logarithmic encoding} \\ 
\cmidrule{2-4} \cmidrule{6-8} 
& \makecell{Nqubits\\ $~$} & \makecell{Npaulis\\ (X, Z)} & \makecell{Density\\ (X, Z)}&
& \makecell{Nqubits\\ $~$} & \makecell{Npaulis\\ (X, Z)} & \makecell{Density\\ (X, Z)} \\
\midrule
\hline %put by GM
General:\\
& $d_S$ & $(4S, d_S - \text{mod}_2 d_S)$
& $(2,2)$ && $\lceil \log_2{d_S} \rceil$ & $(\mathcal{O}({d_S^2}), \mathcal{O}({d_S}))$ &
$(\log_2{d_S}, \log_2{d_S})$\\
\\
\hline %put by GM
Powers of 2:\\
$d_S = 2^n$ & $d_S$ & $(4S, d_S)$
& $(2,2)$ && $\log_2(d_S)$ & $(\sim 1.85^\star d_S, \log_2(d_S) )$ &
$(\log_2(d_S), 1)$\\
\\
\bottomrule
\hline %put by GM
\end{tabular}
\centering
\caption{Resource counts for the linear and logarithmic encoding of spin $S$ systems to qubits. The dimension of the spin system is $d_S = 2S+1$. The first line represents the resource scaling for general $S$. The second line is valid when the dimension $d_S$ of the spin system is a power of 2. The linear dependency with the starred value is determined via a fit to the lowest $10$ powers of 2 ($d_S \leq 1024$) and is valid up to $d_S \sim 3000$. For larger $d_S$ a $log_2{3}$ component becomes relevant according to $n_\text{pauli}(\hat{S}_{x,y}) \approx 1.68 \cdot 10^{-2} d_S^{\log_2{3}} + 1.85 d_S + 1.03$ (c.f. eq.\ref{eq:pauli_Sxy}). Values for spin $Y$ gates are the same as for spin $X$ gates.}
\label{4:tbl:spin_encodings}
\end{table*}

To implement a spin $x$ or $y$ operator, we need to sum over $2S$ terms of the form $XX + YY$ as seen from Eq.~ \ref{4:eq:lin_embedded_splus} and Eq.~\ref{4:eq:lin_spin_xyz}. 
This implies a sum of 
\begin{equation}
    \label{4:eq:lin_pauli_scaling_xy}
    n_\text{pauli}[\hat{S}_{x,y}] = 2 d_S - 2 = 4S
\end{equation}
Pauli strings with support $2$ each.
For the spin $z$ operator, we sum $d_S$ terms of the form $Z$. If $d_S$ is even, i.e. $S$ half integer, then all coefficients $m_s$ in the sum are nonzero and the sum has $d_S$ Pauli terms of density 1. 
On the other hand, if $d_S$ is odd, $S$ integer, then $m_s=0$ is in the spectrum of $\hat{S}_z$ and so one term of the sum in Eq.~\ref{4:eq:lin_spin_xyz} vanishes. 
The result of the linear encoding in this case is a linear combination $d_S-1$ Pauli strings of support $1$. 
In summary we find for the spin $z$ operator
\begin{equation}
    \label{4:eq:lin_pauli_scaling_z}
    n_\text{pauli}[\hat{S}_z] = 
    \begin{cases}
    d_S,~~ &d_S~\text{even}~(S~\text{half integer})\\
    d_S-1,~~ &d_S~\text{odd}~(S~\text{integer})
    \end{cases}
\end{equation}
Note that in contrast to the upper bounds given for the logarithmic encoding, the relations derived above for the linear embedding are exact for all values of the spin $S$.
 
The benefit of the linear encoding is the linear increase %linearly scaling number 
with the system dimension of the number of Pauli strings required to represent any spin operator. 
Moreover these have a constant Pauli support. 
The drawback is that also the number of required qubits grows linearly with the system dimension. 
A summary table to compare the logarithmic and linear encoding is provided in Table~\ref{4:tbl:spin_encodings}.

\subsection{Qubit representation of the fermionic operators}
\label{4:ssec:fermionic_registers}

The procedures to map fermions to qubit are more common and extensively adopted in quantum chemistry applications~\cite{Bravyi2002FermionicComputation}.
They are the \emph{Jordan-Wigner} mapping \cite{Jordan1928UberAquivalenzverbot}, the \emph{parity} mapping, and the \emph{Bravyi-Kitaev} mapping~\cite{Bravyi2002FermionicComputation, Bravyi2017TaperingHamiltonians}. 

All three mappings map fermionic operators on a $N$-fermion register to qubit operators on an $N$-qubit register. 
The sacrifice is that local fermionic interactions are mapped to non-local interactions of $\mathcal{O}({\log N})$ or $\mathcal{O}({N})$. The scaling laws for the three fermionic mappings are summarized in Table~\ref{4:tbl:fermionic_mappings}.

\begin{table}[htb!]\centering
\ra{1}
%\begin{tabular}{@{}lccc@{}}\toprule
\begin{tabular}{ccc}
\hline
\makecell{Mapping} & \makecell{Nqubits} & 
 \makecell{Support} \\ 
\midrule
\hline
Jordan-Wigner & N  & $\mathcal{O}({N})$ \\
Parity & N  & $\mathcal{O}({N})$ \\
Bravyi-Kitaev & N  & $\mathcal{O}({\log{N}})$ \\
\hline
\bottomrule
\end{tabular}
\centering
\caption{ Resource counts for the Jordan-Wigner, Parity and Bravyi-Kitaev mapping of $N$ fermionic modes to qubits. Nqubits is the number of qubits needed to encode $N$ fermionic modes.}
\label{4:tbl:fermionic_mappings}
\end{table}

\section{Resources estimation for the simulation of LQED}
\label{sec:res_qed}

In this section, we will assess the resource requirements for the QLM formulation of $U(1)$ LGT with dynamical Wilson fermions in arbitrary spacial dimension $d$, making use of the results of the previous Sect.~\ref{3:qubits}. 
We will first estimate the number of qubits needed to encode all degrees of freedom in the Hamiltonian of Eq.~\ref{4:eq:H_cqed_rescaled}. 
Secondly, we will discuss the number of Pauli strings required to implement such Hamiltonian.
This information will be translated into a circuit depth estimate in Sect.~\ref{5:real_time}.
All scaling laws will be given in terms of a combination of the model parameters defined in Table~\ref{table:def}.

\begin{table*}
    \begin{tabular}{  l   l }
    \hline
    Variable & Description \\ \hline
    $n_s$  & number of lattice sites \\ 
    {$n_e$\hfill} & number of lattice edges. Scales linearly with $n_s$ in regular lattices. \\
    {$n_p$\hfill} & number of lattice plaquettes. Scales linearly with $n_s$ in regular lattices.\\
    {$n_\text{spinor}$\hfill} & number of spinor components.\\
    {$d$\hfill} & number of spatial lattice dimensions.\\
    {$d_S$\hfill} & dimension of the spin $S$ system in the quantum link model.\\
    {$n_\text{nonzero}(A)$\hfill}& number of nonzero elements of the matrix $A$.\\
    {$n_\text{pauli}[\hat{O}]$\hfill} &total number of Pauli strings in the encoding of the operator $\hat{O}$.\\
   {$n_\text{real}[\hat{O}]$\hfill} &number of Pauli strings with purely real coefficients in the encoding of $\hat{O}$.\\
    {$n_\text{imag}[\hat{O}]$\hfill}& number of Pauli strings with purely imaginary coefficients in the encoding of $\hat{O}$.\\
    {$n_\text{mix}[\hat{O}]$\hfill}& number of Pauli strings with neither purely real nor purely imaginary coefficients \\
    \hline
    \end{tabular}
\caption{ Symbols definition.}
\label{table:def}
\end{table*}

\subsection{Qubit count}
The required number of qubits depends on the number of fermionic and gauge degrees of freedom in the model~\cite{byrnes2006simulating}. 
In the Wilson approach, each physical lattice site hosts $n_\text{spinor}$ fermionic components, so in total we
need to map a fermionic register of length $n_\text{s} n_\text{spinor}$ to qubits. 
This requires $n_\text{s} n_\text{spinor}$ qubits under the previously discussed fermionic mappings.
For the gauge part, we need to map $n_\text{e}$ truncated single link Hilbert spaces to qubits. 
Under the local mappings in Sect.~\ref{4:ssec:spin_registers} this requires  $n_\text{e} \lceil
\log_2{d_S}\rceil$ ($n_\text{e} d_S$) qubits for the logarithmic (linear) spin-to-qubit encoding, respectively.
Examples of necessary sizes of the registers required to simulate prototypical lattices in two and three dimensions are listed in Appendix\ref{app:numbers}.

\subsection{Pauli operators count}
In the following we investigate the number of Pauli strings needed to encode the lattice QED Hamiltonian in Eq.~\ref{4:eq:H_cqed_rescaled} on a quantum register. 
This number plays a fundamental role in determining the run times of various quantum simulation algorithms, such as real-time dynamics, where it affects the time-complexity of the Trotter decomposition, or the Variational Quantum Eigensolver (VQE) method~\cite{peruzzo2014variational}, where instead it determines the number of independent measurements required to estimate the ground state properties.
Notice that in standard quantum chemistry calculations (in the currently adopted second-quantized framework), this number is fixed and scales like $\mathcal{O}(n_\text{qubits}^4)$.

\subsubsection{Mass operator}
The mass term consists of $n_s$ terms of the form $\hbpsi_x \hpsi_x$. 
If we explicitly write out the spinor components this becomes
\begin{equation}
    \label{4:eq:masslike_combo}
    \hbpsi_x \hpsi_x=\sum_{\alpha, \beta} \hpsi_{x,\alpha}^\dagger \gamma^0_{\alpha \beta}
    \hpsi_{x, \beta} \hspace{1cm} \gamma^0_{\alpha \beta} \in \R.
\end{equation}
Therefore we obtain $n_\text{nonzero}(\gamma^0)$ terms of the form $\hpsi_{x,\alpha}^\dagger \hpsi_{x, \beta}$. 
If $\gamma_0$ is diagonal in the chosen representation of the Clifford algebra, then all terms with nonzero coefficients have the form $\hpsi_{x,\alpha}^\dagger \hpsi_{x, \alpha}$, which is mapped to a Pauli string of the form ($...IZI...$$+$$...III...$) under the fermionic mappings in Sect.~\ref{4:ssec:fermionic_registers}. 
Since $\gamma^0$  features a vanishing trace in any matrix representation of the Clifford algebra, the identity terms cancel and we are left with $n_\text{nonzero}(\gamma^0)$ $z$-like terms per site. 
Analogously, if $\gamma^0$ is purely off-diagonal, then we have $n_\text{nonzero}(\gamma^0)/2$ hopping terms ($\hpsi_{x,\alpha}^\dagger \hpsi_{x, \beta} + \hc$), requiring each $2$ Pauli strings according to the  fermion-to-qubit mappings of Sect.~\ref{4:ssec:fermionic_registers}. 
As a result, the number of Pauli strings for the mass term is
\begin{equation}
    \label{4:eq:hmass_pauliscaling}
    n_\text{pauli}[\hat{H}_\text{mass}] = n_s n_\text{nonzero}(\gamma^0)
    \leq n_s n_\text{spinor},
\end{equation}
regardless of whether we have chosen a representation in which $\gamma^0$ is diagonal or not. 

\subsubsection{Hopping and Wilson terms}
The hopping and the Wilson term contain combinations of the form $(\hpsi_x^\dagger \gamma^{\text{mix},k} \hat{U}_{x,k} \hpsi_{x+\hat{k}} + \hc)$, where $\gamma^{\text{mix},k} := \gamma^0 \left(i\gamma^k + r \right)$ is a $d \times d$ matrix.  
Using an explicit representation of the spinor components we have 
\begin{multline}
      \label{4:eq:hoppinglike_combo}
    (\hpsi_x^\dagger \gamma^{\text{mix},k} \hat{U}_{x,k} \hpsi_{x+\hat{k}} + \hc)= \\
    \sum\limits_{\alpha, \beta} (\hpsi_{x,\alpha}^\dagger \gamma^{\text{mix},k}_{\alpha \beta} \hat{U}_{x,k} \hpsi_{x+\hat{k}, \beta} + \hc) \, .
\end{multline}
As this expression is hermitian,  it can only contain purely real Pauli strings. To find their number, we need to count the number of strings with non-vanishing real part in the combination $ \hpsi_{x,\alpha}^\dagger \gamma^{\text{mix},k}_{\alpha \beta} \hat{U}_{x,k} \hpsi_{x+\hat{k}, \beta}$. \\

For the fermionic part, the above terms always operate on two different fermionic sites (due to the different lattice sites $x$ and $x+\hat{k}$). 
The  hopping part therefore always takes the form $(XX+YY+iXY+iYX)$, with two purely real and two purely imaginary Pauli strings for each of the fermionic mappings in  Sect.~\ref{4:ssec:fermionic_registers}.\\ 

For the gauge part, we work for generality with $n_\text{pauli}[\hat{U}] = n_\text{real}[\hat{U}] + n_\text{imag}[\hat{U}] + n_\text{mix}[\hat{U}]$ Pauli strings for the qubit implementation of the truncated gauge field operators $\hat{U}$, inserting later the corresponding values for the two different qubit-mappings of Sect.~\ref{4:ssec:spin_registers}.

The tensor product of the fermionic and gauge parts features then $2(n_\text{real}[\hat{U}] + n_\text{imag}[\hat{U}] + 2n_\text{mix}[\hat{U}])$ Pauli strings with non-vanishing real part. 
After summing over all $n_e(k)$ lattice edges in direction $k$ we find the number of Pauli strings to describe the hopping Hamiltonian
%\begin{align}
\begin{multline}
    \label{4:eq:hhopp_pauliscaling}
    n_\text{pauli}[\hat{H}_\text{hopp}] = 
    \sum\limits_k n_e(k) n_\text{nonzero}(-i\gamma^0\gamma^k)  \times \\
    \times 2
    \left(
    n_\text{real}[\hat{U}] + n_\text{imag}[\hat{U}] + 2n_\text{mix}[\hat{U}]
    \right) \, .
\end{multline}
%\end{align}
We can now insert the specifics of our spin-to-qubit mappings to get upper limits and the scaling behavior for the linear and logarithmic encodings from Sect.\ref{4:ssec:spin_registers}. 
For the logarithmic spin encoding an upper limit is given by 
\begin{align}
    \label{4:eq:hhopp_log_paulibound}
    n_\text{pauli}^\text{log}[\hat{H}_\text{hopp}] \leq 
    16 \, n_e \, n_\text{spinor}^2 d_S^2 \sim \mathcal{O}({n_e n_\text{spinor}^2  d_S^2}) \, .
\end{align}
The scaling improves for the case of a {\it perfectly representable} spin system ($d_S$ a power of 2), 
\begin{align}
    \label{4:eq:hhopp_log_paulibound_perfect}
    n_\text{pauli}^\text{log-perf.}[\hat{H}_\text{hopp}] \leq 
    4 n_e n_\text{spinor}^2 (2.1 d_S) \sim \mathcal{O}({n_e n_\text{spinor}^2  d_S}) \, ,
\end{align}
which is linear in $d_S$. 
For the linear spin encoding instead we find the upper limit
\begin{align}
    n_\text{pauli}^\text{lin}[\hat{H}_\text{hopp}] &= \sum\limits_k n_e(k) n_\text{nonzero}(-i\gamma^0\gamma^k) \, 8 \, 
    (d_S - 1) \\
    &\leq 
    8 n_e n_\text{spinor}^2 (d_S-1) \sim \mathcal{O}({n_e n_\text{spinor}^2  d_S}) \, .
    \label{4:eq:hhopp_lin_paulibound}
\end{align}
The inequality is exact in the case of a square lattice and a representation of the Clifford algebra in which the number of nonzero matrix elements is equal to the number of spinor components.

The above calculations can also be applied to the Wilson correction term  (see Sect~\ref{2:theoryclassical}), which is a combination of mass and hopping like terms. 
As we are interested in the scaling behavior of the total Hamiltonian, we now combine the mass, hopping and Wilson terms into a unique expression that we name $\hat{H}_\text{mhw}$. 
In summary, we find that the number of Pauli strings to implement $\hat{H}_\text{mhw}$ on qubits is
\begin{widetext}
\begin{align}
    n_\text{pauli}[\hat{H}_\text{mhw}] &= n_s n_\text{nonzero}(\gamma^0) + \sum\limits_k n_e(k) n_\text{nonzero}(\gamma^{\text{mix},k }) \, 2 \,  \left(
    n_\text{real}[\hat{U}] + n_\text{imag}[\hat{U}]
    \right) + 2 n_\text{mix}[\hat{U}]\\
    &\leq n_s n_\text{spinor} + n_s d 
    n_\text{spinor}^2  2 \left(
    n_\text{real}[\hat{U}] + n_\text{imag}[\hat{U}] + 2 n_\text{mix}[\hat{U}]
    \right) \\
    &= n_s n_\text{spinor} \left(1 + 2d n_\text{spinor}^2 \left(
    n_\text{real}[\hat{U}] + n_\text{imag}[\hat{U}] + 2 n_\text{mix}[\hat{U}]
    \right) \right) \, .
\end{align}
The final scaling laws for the fermionic sector of the LGT Hamiltonian are therefore
\begin{align}
    \label{4:eq:hmhw_pauliscaling}
    n_\text{pauli}[\hat{H}_\text{mhw}] \sim 
    \begin{cases}
    \mathcal{O}({n_s d n_\text{spinor}^2 d_S^2}),~~ &\tx{logarithmic encoding (general)}.\\
    \mathcal{O}({n_s d n_\text{spinor}^2 d_S^*}),~~ &\tx{logarithmic encoding (perfectly representable)}.\\
   \mathcal{O}({n_s d n_\text{spinor}^2 d_S}),~~ &\tx{linear encoding} \, .
    \end{cases}
\end{align}
\end{widetext}
The starred value is valid in the regime of $d_S \lesssim 3000$. For $d_S$ larger, substitute $d_S \rightarrow d_S^{log_2{3}}\sim d_S^{1.6}$.

\subsubsection{Electric field term}
The electric field Hamiltonian consists of terms of the form $\hat{E}^2$.
In the $U(1)$ quantum link model, the truncated operator $\hat{E}$ is proportional to a $\hat{S}_z$ operator. 
Working in the flux basis i.e., the eigenbasis of $\hat{E}$, all Pauli string in the encoding of a power of $\hat{E}$ can only contain $I$ and $Z$ operators. 
With the logarithmic encoding, this yields the direct loose upper bound
\begin{align}
    \label{4:eq:helec_paulibound_log}
    n_\text{pauli}^\text{log}[\hat{E}^2] \leq 2^{\lceil \log_2{d_S} \rceil} \leq 2 d_S \sim \mathcal{O}({d_S}) \, .
\end{align}
In the case of the linear encoding, we can use Eq.~\ref{4:eq:lin_spin_xyz} to find the exact expression. 
A direct calculation for the cases where $d_S$ is even ($S$ half integer) or odd ($S$ integer) yields, respectively
\begin{align}
    \label{4:eq:helec_paulibound_lin}
    n_\text{pauli}^\text{lin}[\hat{E}^2] =   
    \begin{cases}
    \dfrac{d_S (d_S -1)}{2} + 1,~~ &d_S~\text{even}\\
    \dfrac{(d_S-1)(d_S -2)}{2} + 1,~~ &d_S~\text{odd} \, .
    \end{cases}
\end{align}
Strikingly, the number Pauli strings for the $\hat{S}_z$ operator scales quadratically $\mathcal{O}({d_S^2})$ in the linear encoding as opposed to linearly in the case of the logarithmic encoding. 

When considering the total electronic Hamiltonian (for the whole lattice), we need to implement $n_e$ independent terms of the type $\hat{E}^2$, one for each lattice link.
As a result, the number of Pauli strings for $\hat{H}_\text{elec}$ scales finally as 
\begin{align}
    \label{4:eq:helec_pauliscaling}
    \np(\hat{H}_\tx{elec}) \sim 
    \begin{cases}
    \mathcal{O}({n_e d_S}),~~ &\tx{logarithmic encoding.}\\
    \mathcal{O}({n_e d_S^2}),~~ &\tx{linear encoding.}
    \end{cases}
\end{align}

\subsubsection{Plaquette term}
\label{ss:plaquette}
The plaquette term in the Hamiltonian contains terms of the form $(\hat{U}_\Box + \hat{U}_\Box^\dagger)$ for each plaquette of the lattice. 
%The operators
%$\hat{U}_\Box$ couple four links and are a product of four 
%spin raising and lowering operators $\hat{U}$. $(\hat{U}_\Box + \hat{U}_\Box^\dagger)$
Each operator is hermitian and thus contains purely real Pauli strings only. 
The number of Pauli strings in $(\hat{U}_\Box + \hat{U}_\Box^\dagger)$ is therefore the same as the number of Pauli strings in $\hat{U}_\Box$ with non-zero real part. 
In general
\begin{equation}
    n_\text{pauli}[\hat{U}] = n_\text{real}[\hat{U}] + n_\text{imag}[\hat{U}]
    + n_\text{mix}[\hat{U}] \, ,
\end{equation}
where $n_\text{real}[\hat{U}]$ ($n_\text{imag}[\hat{U}]$ ) denotes the number of purely real (imaginary) strings in the expansion of $\hat{U}$, and $n_\text{mix}[\hat{U}]$ is the number of strings with coefficients that are neither purely real nor purely imaginary.

Since $\hat{U}_\Box = \hat{U}_1 \hat{U}_2 \hat{U}_3^\dagger \hat{U}_4^\dagger$ acts on four different edges, there is no overlap between the supports associated with the different Pauli factors. 
The number of Pauli strings with non-zero real part in the expansion of $\hat{U}_\Box$ is therefore simply  $n_\text{pauli}[\hat{U}]^4$ minus the number of purely imaginary Pauli strings.

Purely imaginary Pauli strings always occur when we multiply an odd number of purely imaginary Pauli strings with a purely real Pauli string. 
For each of the four factors in $\hat{U}_\Box$, the number of such terms is 
\begin{equation}
    4 \, n_\text{real}[\hat{U}] \, n_\text{imag}[\hat{U}]^3
    + 4 \, n_\text{real}[\hat{U}]^3 \, n_\text{imag}[\hat{U}] \, .
\end{equation}

A second way in which purely imaginary terms can occur is associated to the special structure of the coefficients of the mixed Pauli strings. 
Mixed Pauli strings in $\hat{U} \sim (\hat{S}_x +i \hat{S}_y)$ arise from shared terms between the encoded $\hat{S}_x$ and $\hat{S_y}$ operators. 
In particular, they come from the padding of Eq.~\ref{4:eq:spin_in_matrix} when the dimensionality  $d_s$ of the associated QLM spin $S$ system is not a power of two (see Appendix~\ref{app:mixed}).
%\cite{note2}. 
Because the padding is the same for the $\hat{S}_x$ and $\hat{S_y}$ operators, %spin x and y operators, 
the mixed Pauli terms in $\hat{U}$ always have the same coefficient structure $(a+ia)$ with $a$ real. 
Assuming this structure for the 4 coefficients of a plaquette (each characterized by the variables $a,b,c,d$, respectively), the following combinations $(a+ia)(b+ib)~c~d, (a+ia)(b+ib)(ic)~(id), (a+ia)~b~(ic)(d-id) $ also lead to purely imaginary terms.
%\begin{align}
%    (a+ia)(b+ib)~c~~~d~~~~~~ &= \cancelto{0}{(ab-ab)cd} + i(ab+ab)cd \\
%    (a+ia)(b+ib)(ic)~(id)~~ &= \cancelto{0}{(ab-ab)cd} - i(ab+ab)cd\\
%    (a+ia)~b~~~~~(ic)(d-id) &= \cancelto{0}{(ad-ad)bc} + i(ad+ad)bc.
%\end{align}
The number of such terms in the product $\hat{U}_\Box$ amounts to
\begin{equation}
    \begin{aligned}
    n_\text{mix}[\hat{U}]^2 
    \left(
    2 n_\text{real}[\hat{U}]^2 + 2 n_\text{imag}[\hat{U}]^2 + 8 n_\text{imag}[\hat{U}] \, n_\text{real}[\hat{U}]
    \right) \, .
    \end{aligned}
\end{equation}
We therefore obtain 
%From the above it then follows that $\np[\hat{U}_\Box + \hat{U}_\Box^\dagger]$ is
%\begin{equation}
%    \label{4:eq:hplaq:pauli_formula}
%    \begin{aligned}
%    \np[\hat{U}_\Box + \hat{U}_\Box^\dagger] =&
%    n_\text{pauli}(U)^4 - 4 n_\text{real}(U) n_\text{imag}(U)^3
%    - 4 n_\text{imag}(U) n_\text{real}(U)^3  \\
%    &-n_\text{mix}(U)^2 
%    \left[
%    2 n_\text{real}(U)^2 + 2 n_\text{imag}(U)^2 + 8 n_\text{imag}(U)n_\text{real}(U) 
%    \right].
%    \end{aligned}
%\end{equation}
\begin{align}
    \label{4:eq:hplaq:pauli_formula}
    \np&[\hat{U}_\Box + \hat{U}_\Box^\dagger] = 
    n_\text{pauli}[\hat{U}]^4 - 4 \, n_\text{real}[\hat{U}] n_\text{imag}[\hat{U}]^3 - \notag \\
    & 4 \, n_\text{imag}[\hat{U}] n_\text{real}[\hat{U}]^3  - n_\text{mix}[\hat{U}]^2 \times \notag \\
    & \left(
    2 \, n_\text{real}[\hat{U}]^2 + 2 n_\text{imag}[\hat{U}]^2  
     + 8 \,  n_\text{imag}[\hat{U}] \, n_\text{real}[\hat{U}] 
    \right).
\end{align}
This formula is exact for both, the linear and the logarithmic encoding. The total number of Paulis in the plaquette term $\np[\hat{H}_\text{plaq}] = n_\text{plaq}\np[\hat{U}_\Box + \hat{U}_\Box^\dagger]$  scales therefore as 
\begin{equation}
    \label{4:eq:hplaq_pauliscaling}
    \np[\hat{H}_\text{plaq}]  
    %n_\text{plaq}\np[\hat{U}_\Box + \hat{U}_\Box^\dagger] 
    \sim 
    \begin{cases}
    \mathcal{O}({n_s d d_S^8}),~~ &\tx{log. enc. (general)}\\
    \mathcal{O}({n_s d d_S^4}^*),~~ &\tx{log. enc. (perf. rep.)}\\
    \mathcal{O}({n_s d d_S^4}),~~ &\tx{linear enc.}
    \end{cases}
\end{equation}
The starred value is valid in the regime of $d_S \lesssim 3000$. For $d_S$ larger, substitute $d_S^4 \rightarrow d_S^{4 log_2{3}}\sim d_S^{6.3}$.

\subsection{Discussion}
Before moving to the gate count estimation for real-time propagation (Sect.~\ref{5:real_time}), we briefly summarize the results obtained in this Section. 
Concerning the Pauli count, the plaquette term $H_\text{plaq}$ is the most expensive operator, as expected, due to its strong scaling with $d_S$.

The best overall scaling, and thus the lowest number of required Pauli terms, is achieved by the logarithmic encoding for a QLM with a perfectly representable spin $S$ system. 
The apparent downside of this particular encoding is that the eigenvalue $S_z = 0$, which represents vanishing flux through a link, is not contained in the spectrum for $d_S$ a power of two. 
Instead, the lowest absolute flux values for a perfectly representable system are $\pm 1/2$. 
In this case, a zero-flux state can be generated by adding a constant background electric field of $\theta = 1/2$. 
The background field shifts the spectrum of the flux operator $\hat{E}$ by $\theta$. 
Thereby, it effectively turns the $S_z = -1/2$ eigenvalue into $S_z = 0$ in exchange for having one more positive flux state than negative flux states. 
This trade-off does not affect the precision of the calculation adversely for high values of $S$, as states with high flux values carry large energy penalties from the electric field term.
The logarithmic encoding is also more favorable in terms of the number of required qubits. 
We summarize this analysis in Table \ref{4:tbl:hamiltonian_scaling}.
We conclude that LQED simulations beyond the Schwinger model, which is spatially one-dimensional, are much more computationally demanding due to the presence of plaquette terms. 
While the nominal scaling of the number of Pauli strings is linear in the volume, an implicit dependence on the volume is also present in the choice of $d_S$, which should increase with $n_s$ as we want to represent all physically relevant states on the lattice via the QLM approach (cf. Sect.~\ref{ss:qlm}).
Concrete examples of such estimates are listed in Appendix\ref{app:numbers}.

\begin{table}[htb]
\centering
\ra{1.4}
\begin{tabular}{@{}lccc@{}}\toprule
\hline
Term & \multicolumn{3}{c}{Scaling of number of Pauli strings} \\ 
\cmidrule{2-4} 
& \makecell{log. encoding} & \makecell{log. encoding (perfect)} & \makecell{lin. encoding} \\
\midrule
\hline

$\hat{H}_\text{mass}$ & 
${n_\text{s} n_\text{spinor}}$ & 
${n_\text{s} n_\text{spinor}}$ & 
${n_\text{s} n_\text{spinor}}$
\\

$\hat{H}_\text{hopp}$ & 
${n_\text{s} d n_\text{spinor}^2 d_S^2}$ &
${n_\text{s} d n_\text{spinor}^2 d_S^\star}$ &
${n_\text{s} d n_\text{spinor}^2 d_S}$\\

$\hat{H}_\text{wilson}$  & 
${n_\text{s} d n_\text{spinor}^2 d_S^2}$ &
${n_\text{s} d n_\text{spinor}^2 d_S^\star}$ &
${n_\text{s} d n_\text{spinor}^2 d_S}$
\\

$\hat{H}_\text{elec}$ & 
${n_s d d_S} $ &
${n_s d d_S} $ &
${n_\text{s} d d_S^2}$
\\

$\hat{H}_\text{plaq}$ 
& ${n_s d d_S^8}$ 
& ${n_s d d_S^4 { }^\star}$
& ${n_s d d_S^4}$
\\
\hline
\bottomrule
\end{tabular}
\centering
\caption{The scaling relations for the number of Pauli terms for the terms in the lattice QED Hamiltonian of Eq.~\ref{4:eq:H_cqed_rescaled} are shown for different encodings of the truncated gauge operators. These relations do not depend on whether the Jordan-Wigner, Bravyi-Kitaev or Parity mapping is used for the fermions. The starred values ($^\star$) are valid for $d_S \lesssim 3000$. For higher values of $d_S$ the following scaling $ d_S^4{}^\star \rightarrow d_S^{4\log_2{3}} \sim d_S^{6.3}$ holds (see discussions above).} 
\label{4:tbl:hamiltonian_scaling}
\end{table}

We remark that the determination of the scaling in the number of Pauli strings is crucial for both real-time dynamics and VQE algorithms.
In the VQE approach the expectation value of the Hamiltonian needs to be computed as sum of all the Pauli operators expectation values, therefore this scaling readily determines the time-complexity of the algorithm ~\cite{PhysRevA.92.042303,Kandala2017Hardware-efficientMagnets,torlai2019precise}.

\section{The real-time evolution: implementation and gate count}
\label{5:real_time}

In this section, we study the implementation of a time propagation quantum algorithm for LGT. In particular, we will focus on the Trotter algorithm and its mapping to a quantum circuit with the aim of studying the resources requirements in terms of gate counts.

Time propagation is of paramount importance for different physical applications.
The first and  obvious one is to access dynamical properties, such as creation and propagation of particle-antiparticle pairs, which are notoriously hard to calculate with Euclidean based classical methods~\cite{calzetta2008nonequilibrium}.
The second motivation is that time-evolution is an essential building block of the quantum phase estimation (QPE) algorithm.
This, in turn, can be used to extract ground state energies, by measuring the phase acquired by the time-evolved state (provided that $\ket{\phi(0)}$ is not orthogonal to the exact ground state).

Performing real time evolution of quantum systems amounts to simulating the dynamical evolution of a state $\ket{\phi}$ according to the Schr{\"o}dinger equation 
\begin{equation}
    \label{4:eq:SEQ}
    i \hbar \frac{\partial}{\partial t} \ket{\phi} = \hat{H} \ket{\phi} \, .
\end{equation}
For time-independent Hamiltonians, the solution to this equation is simply given by (setting $\hbar \equiv 1$)
\begin{equation}
    \label{4:eq:TISEQ}
    \ket{\phi(t)} = e^{-i\hat{H}t} \ket{\phi(0)} \, .
\end{equation}
The simulation of Eq.~\ref{4:eq:TISEQ} on a quantum computer, requires therefore the implementation of the operator $e^{-i\hat{H}t}$ as a quantum circuit, as well as a method to initialize a given initial state $\ket{\phi(0)}$ in the quantum register.

In general, the exact exponentiation of an arbitrary hermitian operator $\hat{H}$ is impractical (due to the size of the Hilbert space or the number of gate involved), so we resort to approximated methods. 
The Suzuki-Trotter formula~\cite{trotter1959product,suzuki1976generalized}  provides a controllable way of approximating $e^{-i\hat{H}t}$ with a chosen precision $\varepsilon$ in the operator norm. 
To this end, we decompose the hermitian operator $\hat{H}$ as a sum of non-commuting $K$ terms 
\begin{equation}
    \hat{H} = \sum\limits_{k=1}^K \hat{H}_k \, ,
\end{equation}
To propagate the system for a finite time $t$ under the action of the Hamiltonian $\hat{H}$, we use the following  approximation~\cite{suzuki1976generalized}
\begin{equation}
    \label{4:eq:trotter_formula}
    %\exp(-i \hat{H} t) \sim \left[ \prod\limits_{k=1}^m \exp(-i \hat{H}_k \dfrac{t}{p}) \right]^p + \order{m^3 \delta^2 \dfrac{t^2}{p}},
    \exp(-i \hat{H} t) \sim \left[ \prod\limits_{k=1}^K \exp(-i \hat{H}_k \dfrac{t}{p}) \right]^p \, ,
\end{equation}
where $p$ is the number of time slices, $K$ is the number or terms in our decomposition of $\hat{H}$ and $\delta$ is the maximum operator norm $\delta = \max_k \|{\hat{H}_k}\|$ in the decomposition of $\hat{H}$. 
The operator in square brackets is called the Trotter step with step size $\Delta t = t/p$. 
For a given error tolerance $\varepsilon$, we  need $p \sim \mathcal{O}({K^3 \delta^2 t^2 / \varepsilon})$ Trotter steps to approximate $e^{-i\hat{H}t}$ with precision $\varepsilon$ in the operator norm.

Advanced splitting techniques, can improve the overall efficiency of the algorithm~\cite{suzuki1976generalized, Hatano_2005}. 
Another enhancement strategy is to order differently the terms $\hat{H_k}$ in the decomposition for each Trotter steps to mitigate rounding errors, resulting in a higher simulation precision~\cite{Childs_2019,tranter2019ordering}.

The quantum circuits for the implementation of the real-time evolution of hermitean operators, which can be expressed as a sum of Pauli strings, are already well documented in the literature, for instance in quantum chemistry applications (see Ref.~\cite{PhysRevA.92.062318}).
For completeness, we simply provide an illustrative example in Appendix~\ref{app:circuit}.

{\bf Circuit depths.} 
We notice that the LGT Hamiltonian used for time propagation is a sum of local terms. 
In particular, even after applying the Jordan Wigner mapping, the mass, the electric field and the plaquette terms remain local.
As a consequence, the time evolution of all the local commuting operators can be done in \emph{parallel} at each Trotter step, as the terms acting on different \emph{link} commute.
For example, all the terms composing the \emph{electric field} operator can be executed in parallel. 
Similar considerations apply for the \emph{plaquette} term, with the difference that one link is shared between 2 (4) plaquette in 2 (3) dimensions, respectively. 
Since this only depends on the local connectivity of the regular lattices, we conclude that the circuit depth for the simulation of one Trotter step remains \emph{constant} with the system size.
This is in sharp contrast with quantum chemistry simulations, which feature a $\mathcal{O}(M^5)$ depth scaling with the Jordan-Wigner encoding of the $M$ molecular orbitals required for the simulation~\cite{kassal2011simulating}.

The only part of the Hamiltonian that is most affected by the by Jordan-Wigner transformation is the hopping term. In this case, the most unfavorable term to encode in the time evolution circuit would be the hopping between the first and the last enumerated sites in the qubit register (assuming periodic boundary conditions).
This requires a tensor of $\mathcal{O}(n_s)$ operator, therefore a $\mathcal{O}(n_s)$ ladder of CNOT gates.

{\bf Gate counts per Trotter step.} 
Our implementation is based of the so-called \emph{canonical universal set of gates} represented by single-qubit rotations and two-qubits entangling CNOT gates.
This set represents a standard choice, which is available in most state-of-the art quantum computers based on superconducting qubits, while, for instance, it is still not native in ion-trap based quantum computers~\cite{schindler2013quantum}.
Notice that, with the advent of fault-tolerant quantum computation, continuously parametrizable gates, such as single qubit rotations along arbitrary axes, will have to be approximated by a finite sequence of discrete gates, using e.g. T gates and $\pi$/8 gates, to within arbitrarily small errors~\cite{nielsen2002quantum}.

Presently, the hardware implementation of CNOT gates represents the most challenging engineering task, while single-qubit rotations can be realized with much higher fidelities. 
In addition, the number of CNOT gates (simply, CNOT count) limits the single qubit count, which should not exceed a fixed multiple of the CNOT operations.
Therefore, we consider the CNOT gates count as the critical parameter to assess the implementability
of a given algorithm in near term quantum hardware.

After the mapping to the qubit space, any $n$-qubit Hamiltonians $\hat{H} = \sum_{n=1}^{4^n} \lambda_n P_n$ can be simulated efficiently, if only polynomially many Pauli string coefficients $\lambda_n$ are non-zero. 
Without any further circuit optimizations, the number of CNOT gates per Pauli string and per single Trotter step is
\begin{equation}
    \label{4:eq:trotter_CNOT_formula}
    n_\text{CNOT} = 2 \sum\limits_{ \lbrace n | \lambda_n \neq 0 \rbrace } \left( \text{support}(P_n) - 1 \right) \, .
\end{equation}
This estimation assumes an ideal connectivity between all qubits.
However, on physical quantum hardware ideal connectivity is typically not affordable and therefore additional SWAP gates must be included in order to achieve entanglement between any pair of qubits (at the expense of 3 CNOT per SWAP).

In summary, for the proposed lattice QED implementation, we can estimate the required number of CNOT gates per Trotter step using Eq.~ \ref{4:eq:trotter_CNOT_formula}, the scaling for the number of Pauli strings in the Hamiltonian from Sect.~\ref{sec:res_qed} and the scaling for the Pauli support from Sect.~\ref{3:qubits} 
\begin{table*}[htb]
\centering
\ra{1.4}
\begin{tabular}{@{}lcc@{}}\toprule
\hline
Term & \multicolumn{2}{c}{Scaling of the CNOT count per Trotter step} \\ 
\cmidrule{2-3} 
& \makecell{log. encoding} & \makecell{log. encoding (perfect)}\\
\midrule
\hline

$\hat{H}_\text{mass}$ & 
${n_\text{s}^2 n_\text{spinor}^2}$ & 
${n_\text{s}^2 n_\text{spinor}^2} $ 
\\

$\hat{H}_\text{hopp}$ & 
${n_\text{s} d n_\text{spinor}^2 d_S^2 (n_\text{s} n_\text{spinor} + \log{d_S})}$ &
${n_\text{s} d n_\text{spinor}^2 d_S^* (n_\text{s} n_\text{spinor} + \log{d_S})}$ 
\\

$\hat{H}_\text{wilson}$  & 
${n_\text{s} d n_\text{spinor}^2 d_S^2(n_\text{s} n_\text{spinor} + \log{d_S})} $ &
${n_\text{s} d n_\text{spinor}^2 d_S^* (n_\text{s} n_\text{spinor} + \log{d_S})} $
\\

$\hat{H}_\text{elec}$ & 
${n_s d d_S  \log{d_S}} $ &
${n_s d d_S  \log{d_S}} $
\\

$\hat{H}_\text{plaq}$ 
& ${n_s d d_S^8  \log{d_S}}$ &
${n_s d {d_S^4}^* \log{d_S}} $
\\
\hline
\bottomrule
\end{tabular}
\centering
\caption{The table shows scaling relations for the (unoptimized) number of CNOT gates per Trotter step for the terms in the lattice QED Hamiltonian in Eq.~ [\ref{4:eq:H_cqed_rescaled}] for the example of a Jordan-Wigner fermionic encoding and a logarithmic spin encoding. Note that the number of sites $n_s$ scales as $n^d$ for a $d$-dimensional hypercubic lattice with $n$ sites in each direction. The starred values ($^\star$) are valid for $d_S \lesssim 3000$. For higher values of $d_S$ each starred value of $d_S$ is replaced by $d_S^{\log_2{3}}\sim d_S^{1.6}$, and $ d_S^4{}^\star \rightarrow d_S^{4\log_2{3}} \sim d_S^{6.3}$ .}
\label{4:tbl:trotter_CNOT_scaling}
\end{table*}
Adopting the Jordan-Wigner fermion encoding with a worst case Pauli support that scales as $\mathcal{O}({n_{qubits}^\text{fermionic}}) \sim \mathcal{O}({n_s n_\text{spinor}})$, and a logarithmic encoding with a Pauli support of $\mathcal{O}({\log{d_S}})$ per spin, the scaling of the number of Pauli strings in Table.~\ref{4:tbl:hamiltonian_scaling} translates in a scaling for the number of CNOT gates per Trotter step summarized also in Table~\ref{4:tbl:trotter_CNOT_scaling}. 
A similar procedure can be use to establish a relation between the scaling of number of Pauli strings and the one of the CNOT gates for the Bravyi-Kitaev or the Parity mapping in the linear spin encoding.

\section{Time propagation in LGT: Illustrative examples and error estimation}
\label{6:example}

In this chapter, we present some demonstrations of the quantum algorithms for LGT developed in this work, with particular emphasis on the study of time-dependent problems and the selection of the adequate Trotter time step. 
All the results are obtained using a classical simulator of the corresponding quantum circuit. 
For this reason, we will limit our analysis to fairly small-size systems, which can be easily simulated on single CPUs while offering a good demonstration of the potential of quantum algorithms. 
To this end, we developed a LGT software library that will become available in a future release of the open-source \texttt{Qiskit} package of quantum algorithms~\cite{Qiskit}. 
This software package enables both classical simulations of the quantum circuits as well as quantum hardware calculations.
It is worth mentioning that it will be only with the advent of larger quantum computers that the computational advantage of these new approaches will become effective, opening up new opportunities for the study of quantum gauge theories.

\subsection{Bare vacuum decay}

We start with the simulation of the bare vacuum decay due to particle-antiparticle fluctuations on a proof-of-concept system.
\begin{figure}[htb]
    \centering
    \includegraphics[scale=0.6]{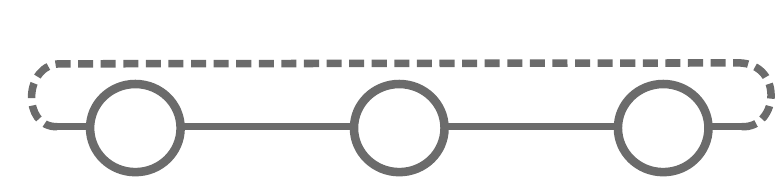}
    \caption{Sketch of the one dimensional spatial lattice with three sites and periodic boundary conditions used as a test case. The circles indicate lattice sites and the connecting lines represent the links of the model.}
    \label{4:fig:vacuum_decay_system}
\end{figure}{}

Let us consider a one-dimensional spatial lattice with three lattice sites and periodic boundary conditions as depicted in~\ref{4:fig:vacuum_decay_system}. 
The Hamiltonian (\ref{4:eq:H_cqed_rescaled}) for this system reduces to
\begin{equation}
\begin{aligned}
    \label{4:eq:vacuum_decay_hamilton}
    \hat{H} &= \sum\limits_{x=1}^3 \dfrac{1}{2a} \left(\hbpsi_x [i \gamma^1 + r]
    \hat{U}_{(x,1)} \hpsi_{x+1} + \hc \right)\\
    &+  \sum\limits_{x=1}^3 \left(m + \dfrac{r}{a} \right) \hbpsi_x \hpsi_x + \dfrac{e^2}{2} \sum\limits_\text{links} \hat{E}_{(x,1)}^2  
\end{aligned}
\end{equation}
where we identified $x=4$ with $x=1$ according to the periodic boundary conditions.
Each lattice site hosts a two-component fermionic field $\hpsi_x$ and therefore requires two qubits. 
We will work in the Dirac representation of the Clifford algebra $Cl(1,1)$ with
\begin{equation}
    \gamma^0 = \sigma^z \hspace{1.5cm} \gamma^1 = i\sigma^x \, .
\end{equation}
Further, we choose the Jordan Wigner fermion-to-qubit mapping and a logarithmic spin-to-qubit encoding. 
With this framework and the Dirac representation of the fermionic degrees of freedom, the computational basis states correspond to the simultaneous eigenstates of the site particle number $\hat{n}_x = \hbpsi_x \hpsi_x$, the site charge $\hat{q}_x = e\hpsi^\dagger_x \hpsi_x$ and the link electric flux $\hat{E}_{(x,1)}$ operators and factor into tensor products of single site and single link states. 
These eigenstates will be called {\it configurations} in the following. 
Therefore, the configurations can be labelled by the set of quantum numbers $n_x, q_x, E_x$ for all lattice sites $x$. 
Within our convention, a state with quantum numbers $n_x=1$ and $q_x=e$ correspond to a \emph{particle} $\ket{p}$ at site $x$, while $n_x=1$ and $q_x=-e$ corresponds to an \emph{anti-particle} $\ket{a}$ at $x$. 
If a site is labelled by quantum numbers $n_x=0$ and $q_x=0$ we have vacuum $\ket{\circ}$ at site $x$, and for quantum numbers $n_x=2$ and $q_x=0$ we speak of a pair-state $\ket{b}$ at site $x$. 
A summary of the notation is given in Table~\ref{5:tbl:physical_interpretation}.

\begin{table}[htb]
\centering
\ra{1.4}
\begin{tabular}{@{}ccl@{}}\toprule
\hline
Site particle number $\hat{n}_x$ & Site charge $\hat{q}_x$ & State at lattice site $x$ \\
\midrule
\hline

$0$ & $0$ & vacuum  $\ket{\circ}$ \\
$1$ & $e$ & particle  $\ket{p}$ \\
$1$ & $-e$ & anti-particle  $\ket{a}$ \\
$2$ & $0$ & pair-state $\ket{b}$
\\
\hline
\bottomrule
\end{tabular}
\centering
\caption{The physical interpretation of the {\it configurations} at a lattice site $x$. The total lattice configurations are product states of the single lattice site configurations, one for each lattice site. }
\label{5:tbl:physical_interpretation}
\end{table}
Concerning the gauge fields, we choose a spin truncation characterized by $S=1$ for the quantum link model, which leads to a non-degenerate ground state; in the logarithmic encoding we need two qubits to represent as single link. 
The chosen (arbitrary) encoding of the single link configurations is given by
\begin{align*}
    \ket{00} &= \ket{\text{unphysical}} \\
    \ket{01} &= \ket{\text{Flux 1}} =: \ket{\rightarrow} \\
    \ket{10} &= \ket{\text{Flux 0}} =: \ket{-} \\
    \ket{11} &= \ket{\text{Flux -1}} =: \ket{\leftarrow} \, .
\end{align*}
With $S=1$ the one dimensional $U(1)$ quantum link model in \ref{4:fig:vacuum_decay_system} has a total of $4^3 \, 3^3=1728$ possible configurations, 48 of which satisfy the Gauss law constraint
\begin{equation}
    \hat{G}_x \ket{\phi} = \left[ \hat{E}_{x-1, 1} - \hat{E}_{x,1} - e \hpsi_x^\dagger \hpsi_x 
    \right] \ket{\phi} = 0 \, , \hspace{0.4cm} \forall x \in \Gamma \, . 
    \label{4:eq:simple_lattice_gauss_law}
\end{equation}
The system is initialized in the bare vacuum state $\ket{\phi_0} = \ket{\circ - \circ - \circ -}$, which corresponds to the ground state of the Hamiltonian in the limit $m \to \infty, e \to \infty$.
In this state, the particle expectation number 
\begin{equation}
    \langle\hat{n}_x \rangle_{\phi_0} = \langle\hbpsi_x \hpsi_x \rangle_{\phi_0}
\end{equation} 
is zero for all sites $x$, and the expectation value of the site charge
\begin{equation}
    \langle \hat{q}_x \rangle_{\phi_0} = e \, \langle \hpsi^\dagger_x \hpsi_x\rangle_{\phi_0}
\end{equation}
is also zero for each lattice site, as well as the flux.
%Additionally, no flux is present in $\ket{\phi_0}$.
However, since $|\phi_0\rangle$ is not an eigenstate of Eq.~\ref{4:eq:vacuum_decay_hamilton} at finite $m$, it undergoes a non-trivial time evolution.

With the adopted encoding, the Hamiltonian contains 466 Pauli strings, resulting into a number of 3302 CNOT gates to perform a single Trotter step (see Eq.~\ref{4:eq:trotter_CNOT_formula}). 
This value can likely be reduced by using transpilers for automatic or manual circuit optimization~\cite{hner2018using, H_ner_2018, Nam_2018}.
We also noticed that the choice of the fermion-to-qubit mapping changes only slightly such estimate, as the gate counts obtained with the Bravyi-Kitaev and the Parity mapping reads 3434 and 3178, respectively. 
 
\begin{figure}[htb]
    \centering
    \includegraphics[width=\columnwidth]{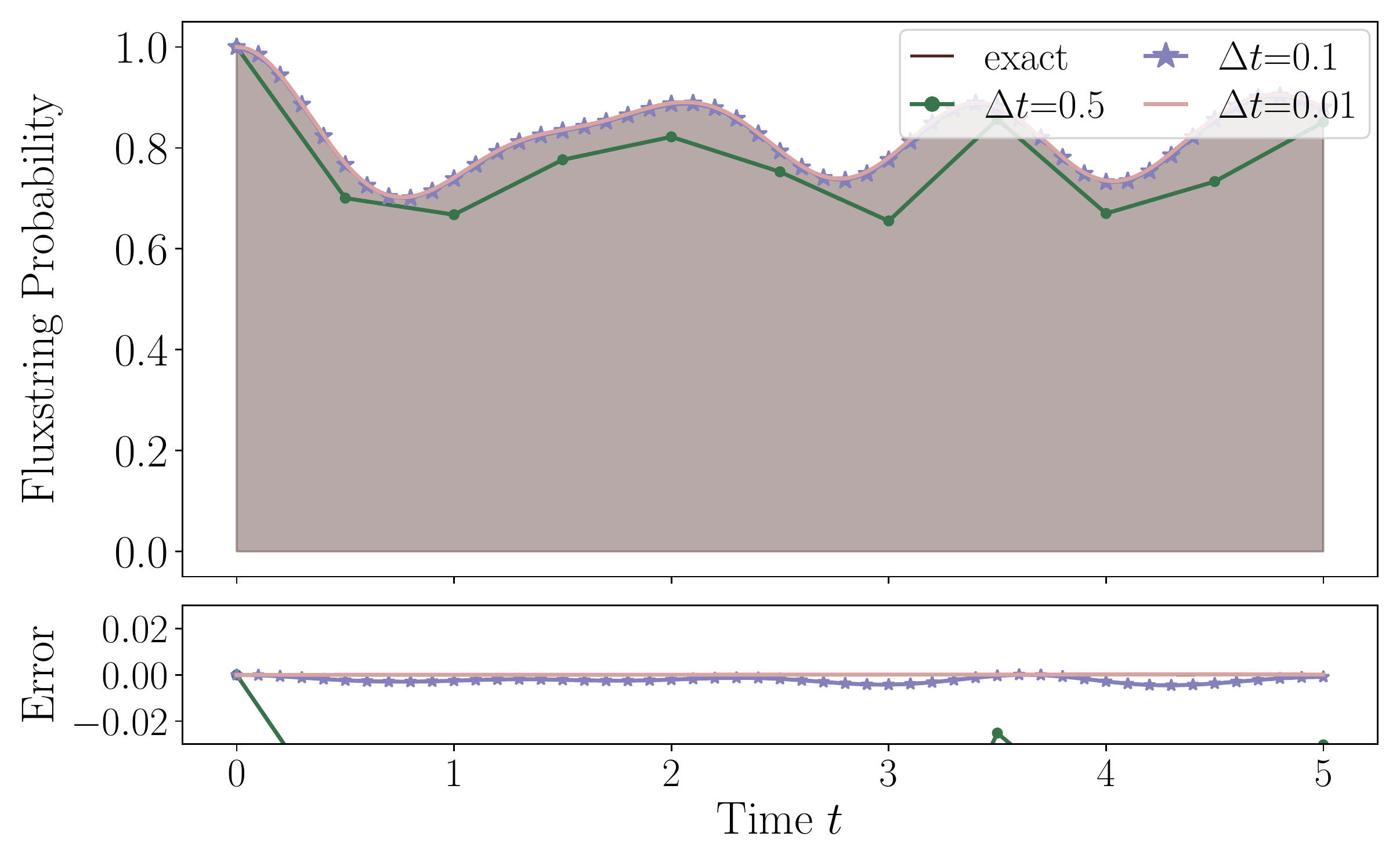}
    \caption{The time evolution of the square of the vacuum persistence probability $|G(t)|^2 =\langle \phi_0 | e^{-iHt}|\phi_0 \rangle$ is plotted in the upper panel for three different Trotter step sizes, $\Delta t$.  For comparison, the exact solution is shown in blue. The lower panel shows the relative Trotter errors for the three simulations.}
    \label{4:fig:bare_vacuum_decay_trotterplot}
\end{figure}

The vacuum persistence amplitude $G(t) = \langle  \phi_0| e^{-i\hat{H}t}| \phi_0 \rangle$ quantifies this decay of the unstable bare vacuum.
The associated probability $|G(t)|^2$ is known as Loschmidt echo~\cite{Monz2016Real-timeComputer} and is shown in Fig.~ \ref{4:fig:bare_vacuum_decay_trotterplot} 
for three different Trotter step values $\Delta t$, and with the model parameters $m=0.5$, $r=1$, $a=0.5$ and $e=\sqrt{2}$.
{}

\begin{figure}[htb]
    \centering
    \includegraphics[width=\columnwidth]{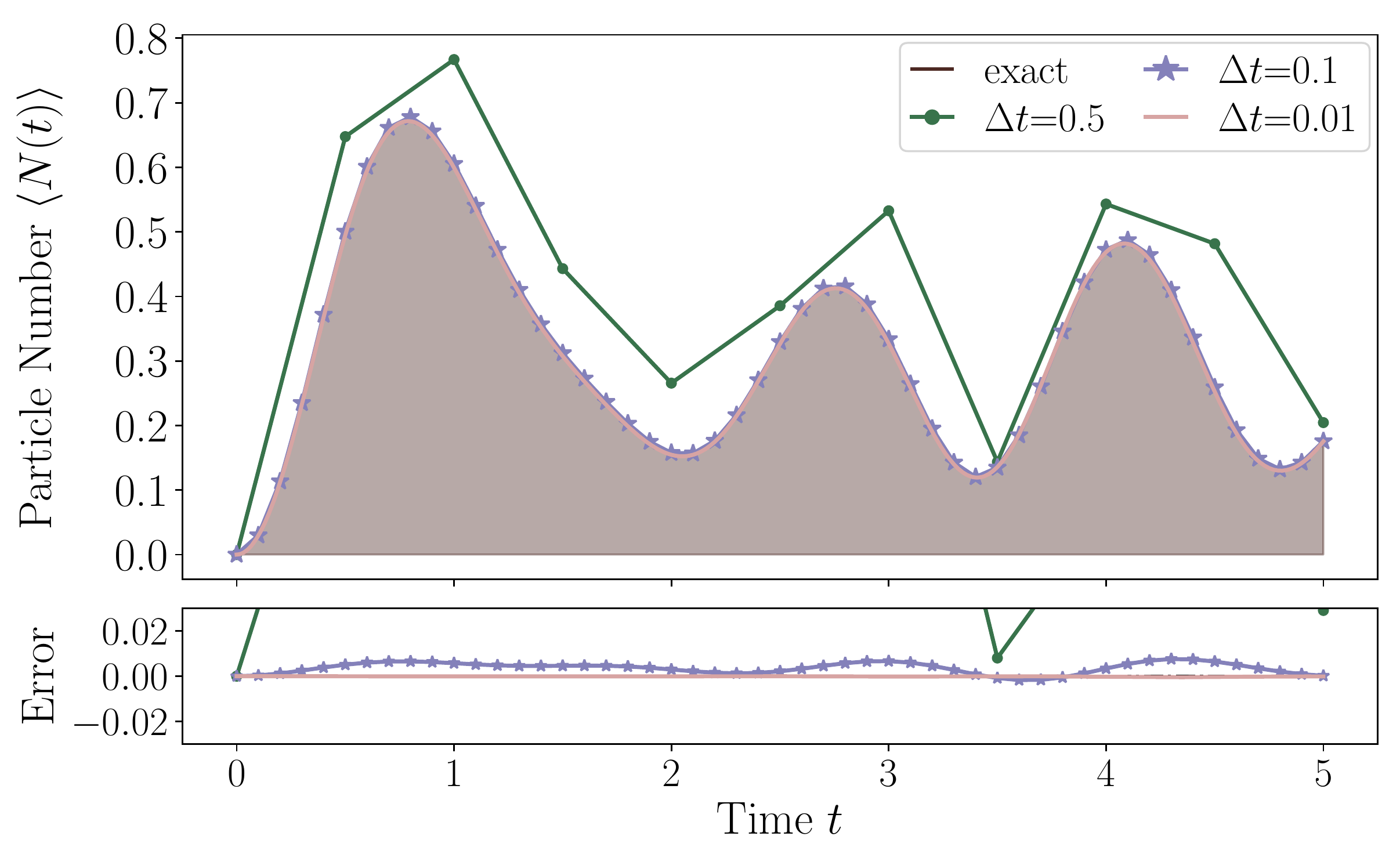}
    \caption{Time evolution of the total particle number $\langle{N(t)}\rangle= \sum_x \langle {\hbpsi_x \hpsi_x} \rangle$ a evaluated for three different Trotter step sizes. The exact solution is included in blue as a reference. The lower panel shows the relative error as a function of the time step used in the simulations.
    }
    \label{4:fig:bare_vacuum_decay_particle_number}
\end{figure}

To better understand the nature of this process, we also monitored 
the expectation value of the particle number operator (shown in Fig.~\ref{4:fig:bare_vacuum_decay_particle_number}). 
We notice that the bare vacuum decay consists in an initial phase of rapid pair creation, which is followed by several recombinations, as pair-creation and annihilation effects compete.

By measuring the time-evolving state in the computational basis at intermediate times we can gain insights into the decomposition of the state into lattice configurations. 

\begin{figure}[htb]
    \centering
    \includegraphics[width=\columnwidth]{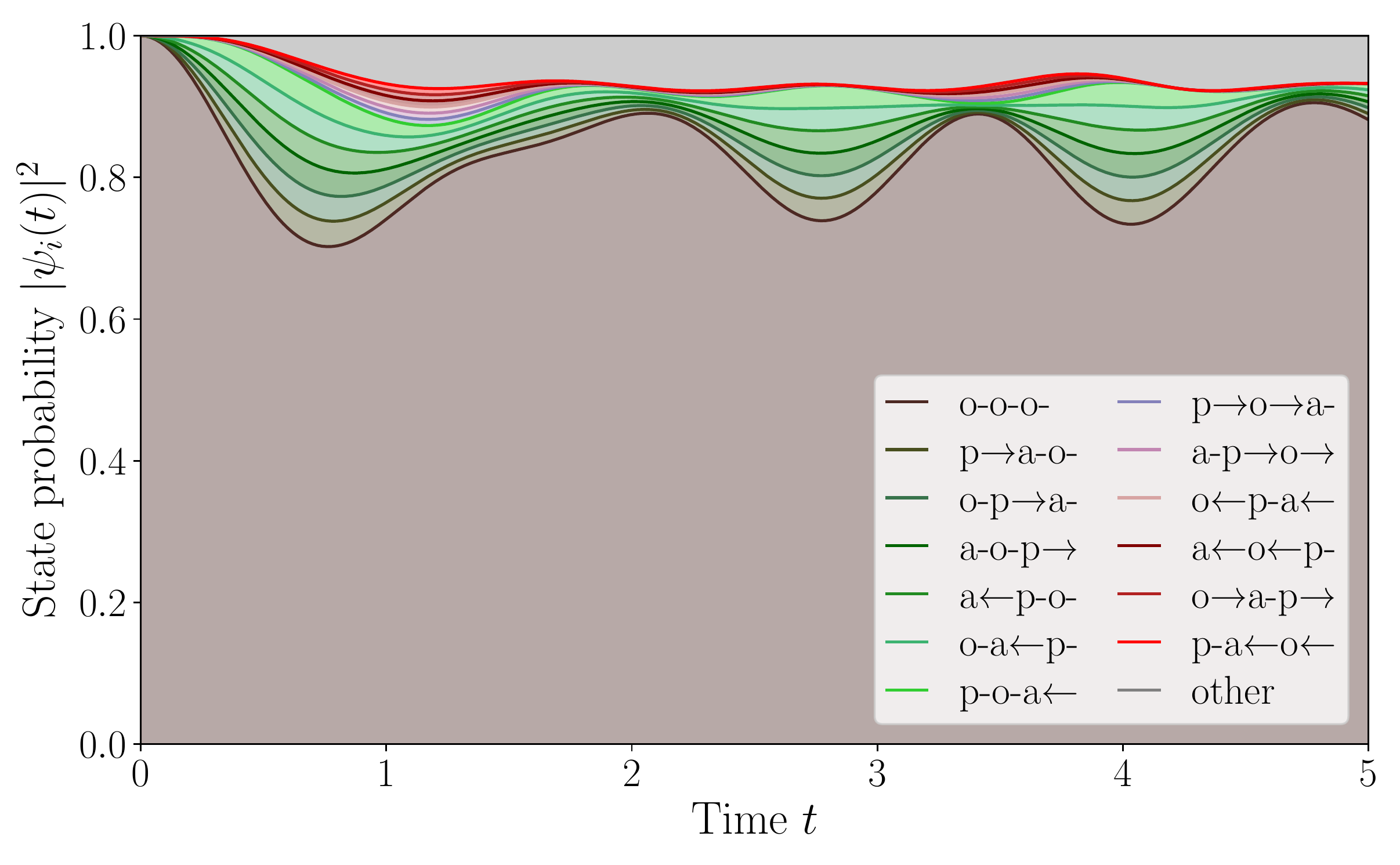}
    \caption{
    Time evolution of the string state projected onto the one-dimensional lattice configurations reported in the inset.
    Single pair configurations with total flux equal to 1 (one excited link) are shown in a green color palette while single pair configurations with a total flux equal of 2 (two excited links) are shown in a red color palette. Higher order excitations are grouped into \textit{other}.}
    \label{4:fig:bare_vacuum_decay_probabilities}
\end{figure}

%\iffalse
As an example, at time $t=0.4$ we obtain
\begin{equation}
    \ket{\phi_0 (t=0.4)} 
    \begin{cases}
        82.5\%~~ &\ket{\circ -~\circ - \circ -}\\
        ~2.7\%~~ &\ket{p \rightarrow a - \circ -}\\
        ~2.7\%~~ &\ket{\circ -p \rightarrow a -}\\
        ~2.7\%~~ &\ket{a - \circ - p \rightarrow} \\
        ~2.7\%~~ &\ket{a \leftarrow p - \circ -}\\
        ~2.7\%~~ &\ket{\circ - a \leftarrow p -}\\
        ~2.7\%~~ &\ket{p - \circ - a \leftarrow} \\
        ~\text{rest}     &\text{others} \,  \\
    \end{cases}
\end{equation}
%\fi
where $\circ, p, a$ indicate, respectively, vacuum, particle or antiparticle occupancy of a site, and $-, \rightarrow, \leftarrow$ the absence or the presence of a flux line in a given direction.
A more in depth picture of how the bare vacuum decays is given in Fig.~ \ref{4:fig:bare_vacuum_decay_probabilities}, where the probability of generating different \emph{configurations} are illustrated using a color code. 
{}

Single pair configurations with absolute total flux equal to $1$, i.e. one excited link, are shown in green scale while single pair configurations with an absolute total flux equal of $2$, i.e. two excited links, are shown in red scale. 
Higher order excitations are labelled as {\it other}. 
Fig.~\ref{4:fig:bare_vacuum_decay_probabilities} nicely illustrates how the initial pair states are generated locally by the hopping term of the Hamiltonian in 
Eq.~\ref{4:eq:vacuum_decay_hamilton} with one unit of flux associated the link in-between. 
The particle and anti-particle pair move then along the lattice, leading to states with higher total flux and states where particles and anti-particles sit at the same lattice sites. 
We also observe how the pairs recombine again back into a vacuum state producing the characteristic vacuum fluctuations. 
%The  phenomenon of vacuum fluctuations and the decay of the bare vacuum are not particular to quantum electrodynamics, but occur in many interacting quantum field theories.

It is worth mentioning that, in a real-hardware setup it is fairly straightforward to extract the quantity $|G(t)|^2$, as well as the other configurations probabilities after every evolution of time $t$, by a simple counting of the measurement readouts in the computational basis.

\subsection{Flux string breaking in one-dimension}
\label{4:ssec:string_breaking}
In this section we study the mechanism of string breaking in LQED. 
%As is well known from classical physics, Electrodynamics is confining in one dimension.
The effects of confinement and string-breaking in one-dimensional QED are reminiscent of the confinement and hadronization in QCD.
String breaking occurs when a particle and an antiparticle are moved further apart, stretching the flux line connecting them according to the Gauss law.
When the energy stored in the flux-string (proportional to its length) exceeds the energy needed to create a particle-antiparticle pair, this can form and the flux-string breaks. 

\begin{figure}[htb]
    \centering
    \includegraphics[scale=0.4]{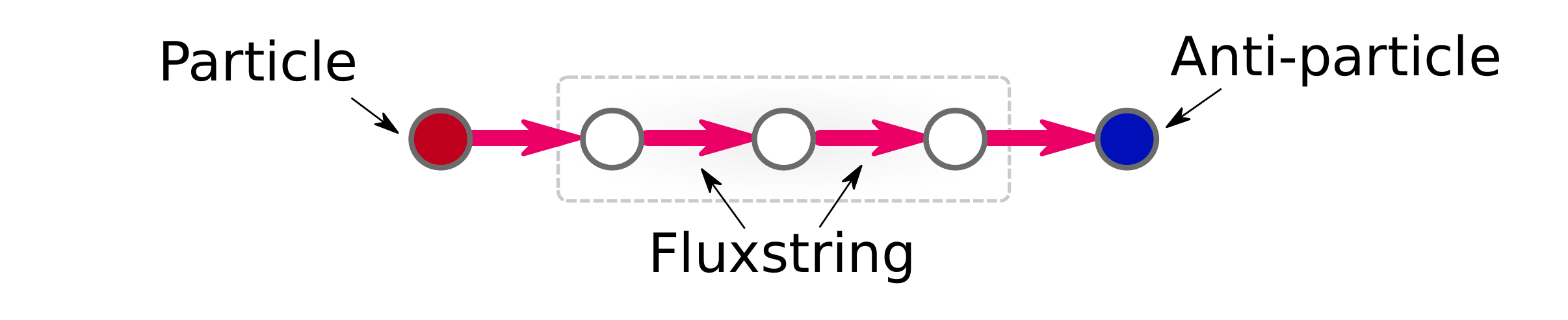}
    \caption{String breaking setup. A one-dimensional lattice with three sites (inside the box) and boundary conditions featuring a particle(antiparticle) to the left(right) of the simulation cell.}
    \label{4:fig:string_breaking_system}
\end{figure}

To study this phenomenon, we look at a one-dimensional lattice with three sites and open boundary conditions as illustrated in Fig.~\ref{4:fig:string_breaking_system}. 
We will again work in the Dirac representation with the Jordan-Wigner fermion-to-qubit mapping and a quantum link model spin truncation of $S=1$ with the logarithmic spin-to-qubit encoding.

The LQED Hamiltonian with the Gauss law correction terms features 305 Pauli strings, leading to an unoptimized CNOT count of 1832 per Trotter step.
The boundary conditions are chosen with a particle on the left hand side and an anti-particle on the right hand side of the system, i.e. the flux values of the links at the boundaries are set to $1$. 
With these boundary conditions, the system features $4^3 3^2=576$ possible configurations, $14$ of which are gauge invariant and satisfy the Gauss law. 
As before, each lattice site is represented by two qubits for the two fermionic components and each link is represented by two qubits in which the spin $S=1$ system is embedded. 

The starting point of our simulation is the initial flux-string state $|\psi \rangle= | \circ \rightarrow \circ \rightarrow \circ \rangle$. 
\iffalse
\begin{figure}[htb]
    \centering
    \includegraphics[width=\columnwidth]{string_breaking_trotterplot_mlarge.pdf}
    \caption{The string probability $P_s(t) = \abs{\mel{\psi}{e^{-iHt}}{\psi}}$ as a function of time in the parameter regime $m \gg e$ for different Trotter step sizes. The model parameters are $m=10$, $e=2$, $a=0.4$, $r=1$, so $m=5e$. We see that the exact solution (blue) is recovered by the circuit simulation for sufficiently 
    small Trotter step size. In this regime, there is not enough energy in the flux-string of the given length to create a particle-antiparticle pair and as a result
    the string does not break. This is seen by the continuously high flux-string probability of almost 1.}
    \label{4:fig:string_breaking_trotterplot_mlarge}
\end{figure}{}
\fi

To illustrate the phenomenon of string breaking and its dependence on the model parameters, we performed the simulation for two different parameter regimes, namely we fix $e=2$, $r=1$ and $a=0.4$, and investigate the dynamics for the cases of $m=5e$ and $m=e/5$. 
In the first setting, the mass was chosen much larger than the field energy, $m \gg e$ ($m = 5e$), such that the energy in the flux-string configuration is stable. % since the energy cost for the generation of a particle anti-particle pair is too high.

\begin{figure}[htb]
    \centering
    \includegraphics[width=\columnwidth]{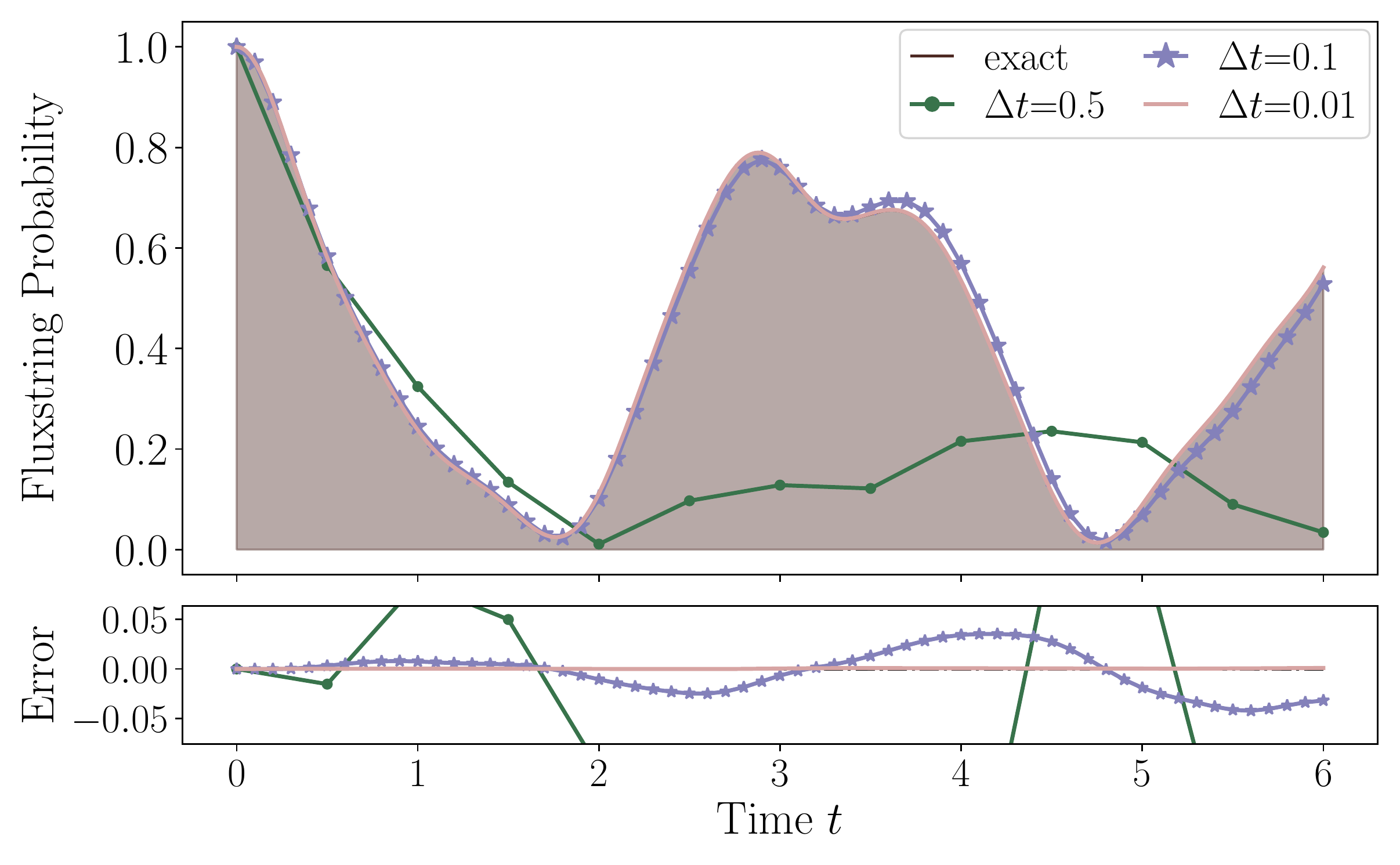}
    \caption{Survival probability for the initial string configuration $P_s(t) = | \langle {\psi}|{e^{-iHt}}|{\psi} \rangle |^2$ as a function of time in the parameter regime $m \ll e$ for different Trotter step sizes. The model parameters are $m=0.4$, $e=2$, $a=0.4$, $r=1$, so $m=e/5$. The exact solution (blue line) is recovered for sufficiently small time steps ($\Delta t < 0.1$).}
    \label{4:fig:string_breaking_trotterplot}
\end{figure}

\begin{figure}[htb]
    \centering
    \includegraphics[width=\columnwidth]{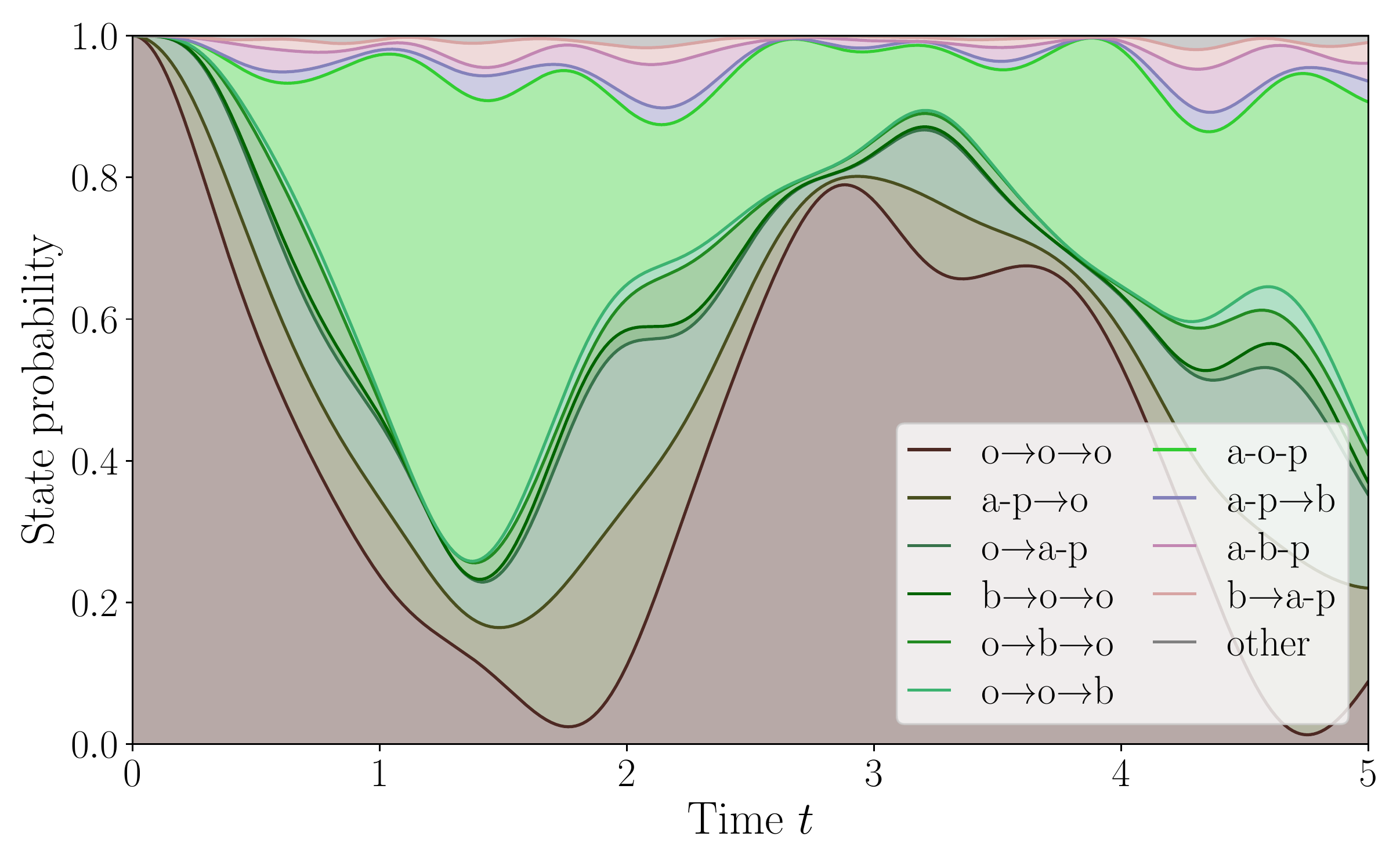}
    \caption{Time evolution of the string state projected onto the lattice configurations reported in the inset. The model parameters are $m=0.4$, $e=2$, $a=0.4$, $r=1$, so $m=e/5$. In this diagram the single pair excitations are plotted with a green color palette, while a red color palette was used for two pair excitations. Higher excitation states are summarized in the category {\it other}.}
    \label{4:fig:string_breaking_probabilities}
\end{figure}

On the other hand, in the regime where $m \ll e$ ($m=e/5$), there is enough energy in the system to create a pair state and break the flux-string (see  Fig.~\ref{4:fig:string_breaking_trotterplot}). 
As for the previous application, we measure the time-evolved flux-string state in the computational basis for various times $t$ to find the probability flow diagram in Fig.~\ref{4:fig:string_breaking_probabilities}. 
The string breaks initially by the local action of the hopping term in the Hamiltonian in Eq.~\ref{4:eq:vacuum_decay_hamilton} (around $t\approx0.2$), then generated particle and antiparticle diffuse to the boundaries, to shield the system from the external flux ($t\approx1.4$). 
As time progresses further, the system oscillates between predominantly flux-like and particle-like configurations.
The phenomenon of string breaking also occurs in QCD, where it is known as \emph{hadronization}. 
As quark and an anti-quark separate, a string of color electric flux between them is formed. 
In the same way as for one-dimensional QED, this color electric flux-string can break by pair creation of quark and anti-quark, forming a meson at each edge of the former flux-string.

\subsection{Double plaquette system in two-dimensions}
\begin{figure}[htb]
    \centering
    \includegraphics[width=0.5\columnwidth]{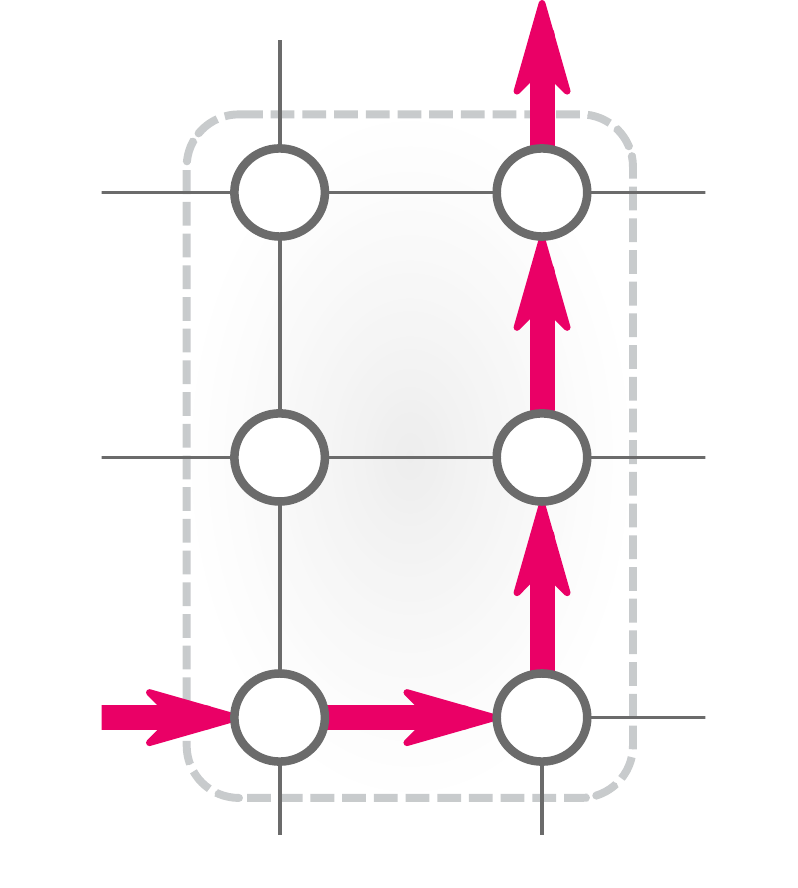}
    \caption{The two dimensional model with six fermionic sites (circles) and links between them. We use white circles to illustrate the vacuum state. The red arrows in the figure illustrate the initial fluxstring through the system.}
    \label{fig:6sites}
\end{figure}

To demonstrate the applicability of our framework for $U(1)$ LGTs in arbitrary dimensions, we represent an example of a genuine 2D system, which features a 6 site lattice arranged as shown in Fig.~\ref{fig:6sites}.
The Hamiltonian describing this system includes the \emph{plaquette} term (see Eq.~\ref{4:eq:H_cqed_rescaled}), therefore the gauge field  degrees of freedom cannot be integrated out, as in one-dimensional example of Ref.~\cite{Kokail2018Self-VerifyingModel}.
As for the one-dimension system of the previous section, also in this case we study the non-equilibrium phenomenon of flux-string breaking, by imposing the boundary condition as in Fig.~\ref{fig:6sites}.
In the same figure, we also show the chosen initial flux-string state configuration. 
The boxed region is simulated dynamically while edges outside of the box are static boundary conditions. 

It is important to note that there exist several configurations realizing a flux-string, depending on the \emph{path} traversed by flux. 
The one selected as initial condition for our simulation is also reported in Fig.~\ref{fig:6sites}.
White circles correspond to a vacuum at a given site and red arrows correspond to one unit of flux along the given edge. 
We will use a blue circle to represent an anti-particle and a red circle to represent a particle at the given site, respectively. 
An arrow on a link signifies one unit of electric flux $ E = (0.5 + \theta)e = e $ in the direction of the arrowhead, while no arrow signifies vanishing flux $ E = (-0.5 + \theta)e = 0 $ through that link. 

In all simulations we use the logarithmic spin-to-qubit encoding with a perfectly representable spin truncation of $S=1/2$.
With this setting a register of $19$ qubits is required to store the state, as the system features $6$ sites and $7$ links.
We added a non-trivial background electric field of $\theta = 0.5$ along the positive $x$ and $y$ axes to generate a zero flux mode in the spectrum. 
The chosen model parameters are $m=0.4$, $e=2$, $a=0.4$, $r=1$, so $m=e/5$. 

\begin{figure}[htb]
    \centering
    \includegraphics[width=\columnwidth]{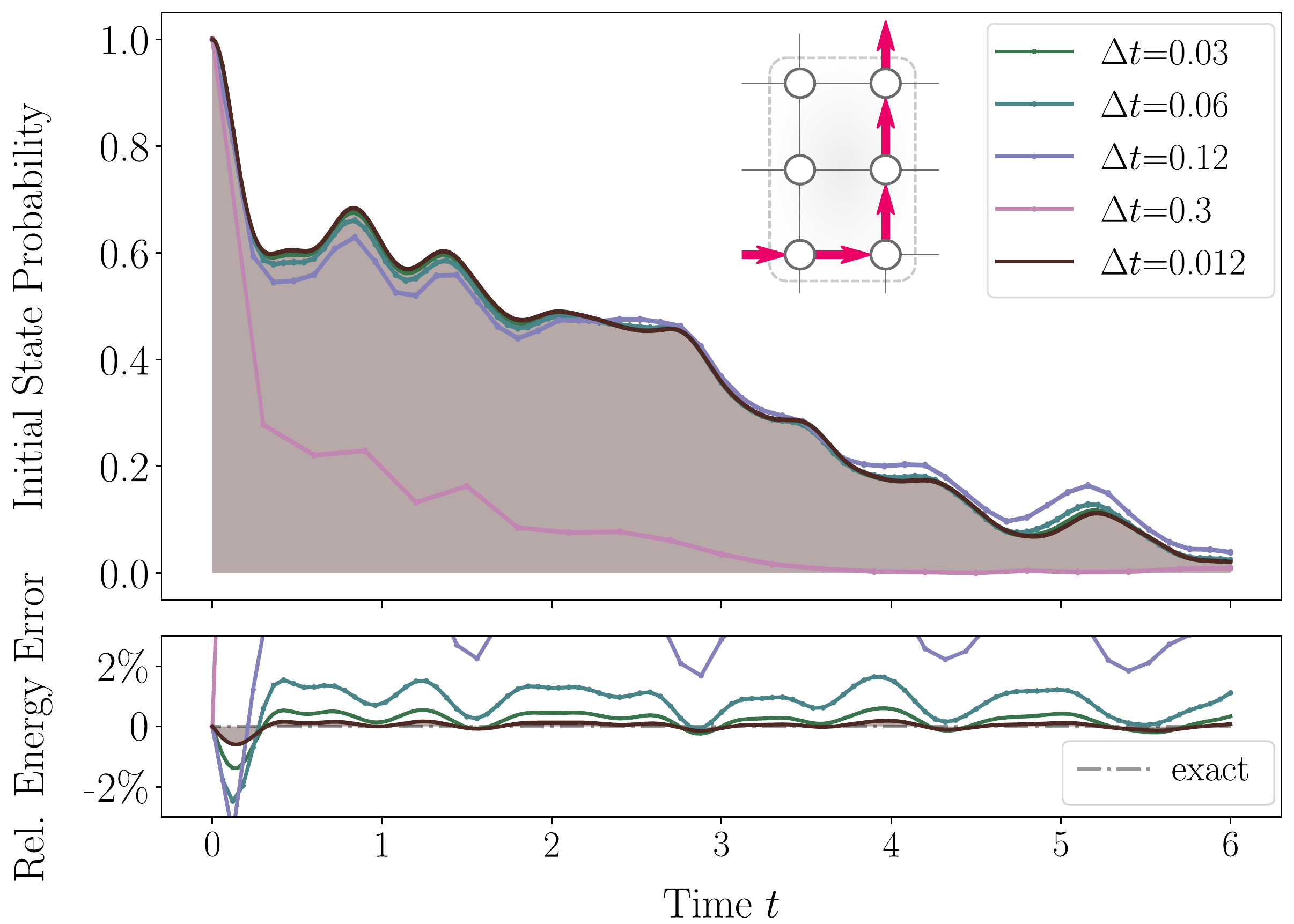}
    \caption{ The upper panel shows the probability $G(t) = | \langle \psi | e^{-iHt} | \psi \rangle |^2$ of finding the double plaquette system in the initial flux-string state shown in the inset. 
    The five curves correspond to simulations performed with five different Trotter timesteps $\Delta_t$.
    The lower panel shows the relative energy error $ \langle \psi(t) | (\hat{H}-E_0)/E_0 | \psi(t) \rangle $ associated to the different curves of the upper panel.
    }
    \label{fig:double_plaquette_trotter}
\end{figure}

\begin{figure}[htb]
    \centering
    \includegraphics[width=\columnwidth]{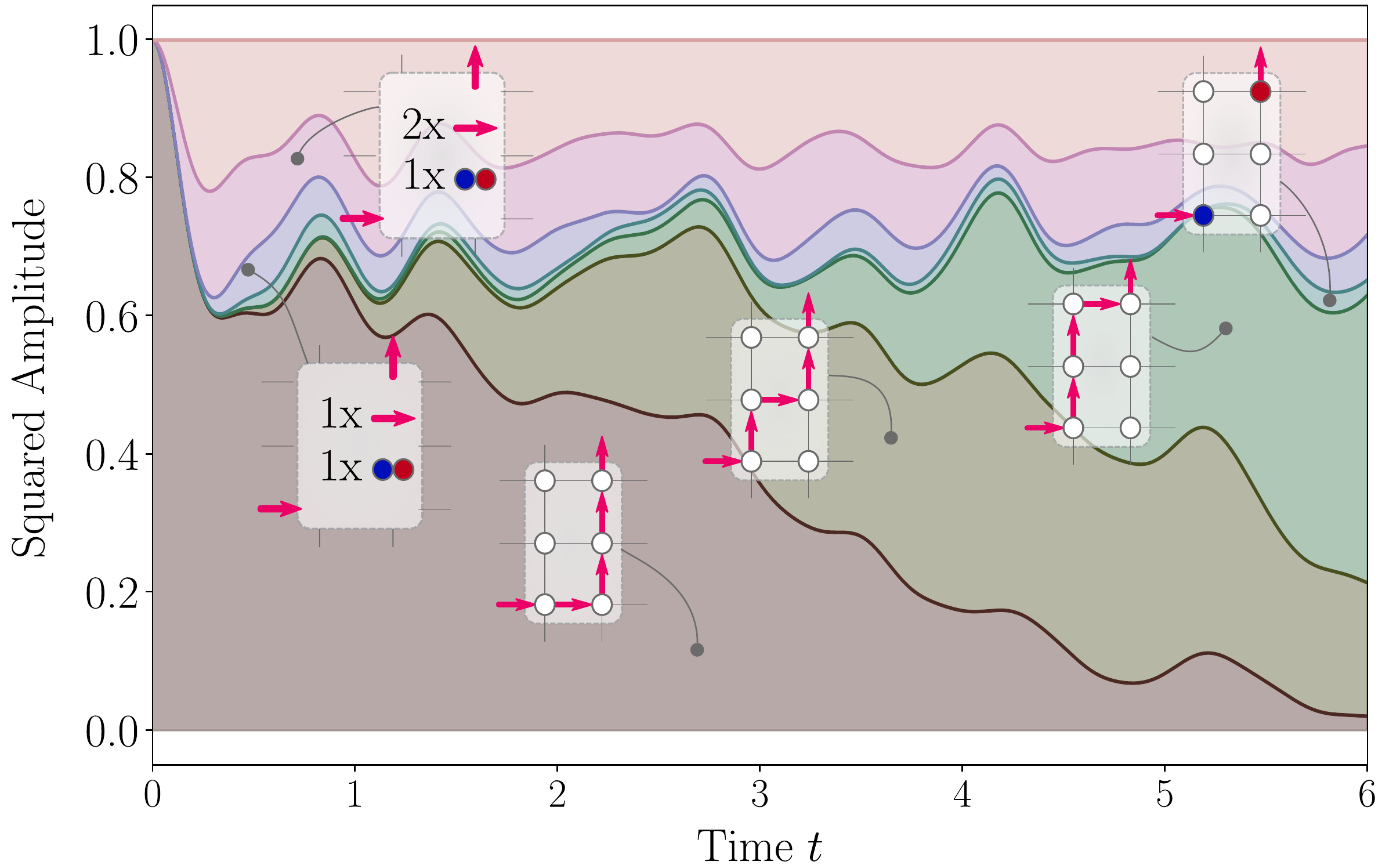}
    \caption{
    The time evolution of the probabilities for the lattice configurations shown in the different insets, obtained propagating the initial configuration (see Fig.~\ref{fig:6sites}) with a Trotter step of $\Delta_t=0.012$.
    The bottom-most curve is the residual probability of observing the initial flux-string state. The next two curves correspond to the probability of observing a state where the electric field lines rearrange along the links shown in the corresponding insets. The displacement of the flux-string is induced by plaquette term. The fourth curve from the bottom corresponds to the probability of observing a broken string configuration with optimal shield of the external flux. The fifth (sixth) curve is the combined probability of finding the system in a configuration with one pair and one (two) links in an excited flux state. Lastly, the topmost curve combines the probabilities of finding the system in any other gauge conserving configuration with a higher number of flux or pair excitations.}
    \label{fig:double_plaquette_probabilities}
\end{figure}

The time-evolution of the initial state is shown in Fig.~\ref{fig:double_plaquette_trotter}, where we depict the decay of the flux-string state probability $G(t)$ decays over time.
The results are given for different Trotter timesteps $\Delta_t$ and show convergence for values $\Delta_t \sim 0.01$.
Fig.~\ref{fig:double_plaquette_probabilities} shows the probability flow diagram with the site and link configurations sampled during the dynamics.
%Let us again inspect a probability flow diagram to analyze the dynamics with which the initial flux-string state breaks. 
%Two types of dynamics occur in the two-dimensional lattice system: 
In this two-dimensional lattice model, we can observe two different dynamical effects: 
(1) Firstly, the flux-string can break via the creation of a pair state, analogous to the one-dimensional case. 
This dynamics is induced by the action of the hopping term. 
(2) Secondly, the location of the flux-string can oscillate in space. 
This second dynamical behaviour is generated by the plaquette term in the Hamiltonina and does not have an analog in one dimension. 
%It corresponds to fluctuations in the location of the flux-string. 
The relative speed of these dynamics is determined by the strength of the corresponding terms in the Hamiltonian.
Even though very simple, this two-dimensional model can already give a glimpse into the richness of the dynamics in gauge field theories.

\subsection{Resources estimation for larger-scale simulations}

We end this Section by reporting resources estimation to perform real-time evolution on systems which are state-of-the-art for tensor-network simulations and beyond.
Ref~\cite{felser2019two} reports a study of the finite density phase diagram in the $S=1$ QLM representation on two-dimensional lattice sizes of up to $16 \times 16$ sites, in the staggered fermions encoding, and with advanced tensor-network classical simulations. 
In proposed Wilson fermions representation, this setup would correspond to a $4 \times 4$ physical site lattice.
To encode the wavefunction in the $S=1$ case, a total register of $80$ qubits would be required: $32$ to encode the matter fields and $48$ for the gauge fields (cfn. Appendix~\ref{app:numbers}). 
The same resources will allow the simulation with a larger truncation value of $S=3/2$.

In Ref~\cite{felser2019two}, it is shown that a truncation corresponding to $S=1$ is sufficient to provide a satisfactory accuracy for computing the ground state.
However, for accurate real-time dynamics simulations a larger truncation of the QLM is expected to be necessary.
The optimal truncation value cannot only be determined \emph{a-priori}.
Here, we only stress the fact that for this system ($4 \times 4$ plaquette) an exact representation could be still achieved with $S=8$.
In fact, the maximum flux traversing a single link can happen when particles and anti-particles are maximally separated in the lattice, and when a flux string traverses first the 8 particle in a \emph{zig-zag} path and then the 8 anti-particle in a similar fashion.

Here we report also the number of Pauli operators needed to represent the Hamiltonian used for time propagation.
Assuming that $S=1$ is a satisfactory value for the study of the dynamics around the prepared ground-state, then the total number of terms in the Hamiltonian is $ \sim 2.5 \times 10^5$.
Interestingly, this number reduces to $\sim 1.2 \times 10^4$ in the case of the perfectly representable setting with $S=3/2$ under periodic boundary conditions.
A further reduction ($\sim 7.7 \times 10^3$) is obtained when open boundary conditions are applied.
The number of Pauli terms for different regular lattices can be reconstructed using the information in Tables of Appendix~\ref{app:mixed}.
For example, the simulation of a cubic lattice of $100 \times 100 \times 100$ sites would require a register of $\mathcal(1-2 \times 10^7$ qubits for QLM truncation values $S$ between $1$ and $31$.
On the other hand, assuming a reasonable truncation value of $S=31$ (that is, perfectly representable) a single plaquette term would require the encoding of $4.2 \times 10^7$ Pauli operators to perform time evolution.
%As discussed, this is the dominant contribution to the requirement of entangling gate resources.
However, most of the single plaquette operators can be \emph{executed} in parallel resulting in constant circuit depth for increasing system size, but, nonetheless, a rapidly increasing depth as a function of the QLM truncation value $S$ (cfn. Sect.~\ref{5:real_time}).

\section{Conclusions}
\label{7:conclusion}

In this work, we present a thorough analysis of the implementation of lattice gauge theories (LGT) in the framework of quantum computing putting particular emphases on the extension to different gauge groups other than $U(1)$ and the scalability to arbitrary spatial dimensions.
To this end, we used a Hamiltonian formulation of LGT in discretized space coordinates with continuous time variable and performed a 
\emph{detailed} resource count estimates for future implementations in universal quantum computers equipped with a \emph{universal canonical gates set}.
%\emph{detailed} resource count estimates to implement lattice gauge theories defined over arbitrary spatial dimensions on a universal quantum computer equipped with the \emph{canonical universal set of gates}.
As a demonstration, we describe step-by-step the implementation of lattice QED, prototypical example of an $U(1)$ lattice gauge theory.
Here we focus on three main aspects, (\emph{i.}) the scaling of the required qubit register, (\emph{ii.}) the scaling of the required number of Pauli terms in the Hamiltonian, and (\emph{iii.}) the scaling of the number of CNOT gates to implement real-time evolution.

Concerning the qubit resources, the gauge-fields represent the most costly dynamical variable to encode.
Therefore, we adopt the Wilson fermion approach as it will optimally reduce the total number of links to be simulated. 
This choice (which is novel in this field), takes on greater importance in large dimensionality, and in particular in view of future QCD applications, where the number of fermion components increases to six.

We discuss three fermion-to-qubit mappings to represent the fermionic fields in the qubit register, as well as two field-to-qubit mapping to include the gauge fields as dynamical variables following the quantum link model approach.
We find that the qubits resource can be made linearly scaling with the volume by introducing a systematic truncation of the flux traversing the single edges.
Physically this means imposing a cut-off on the amplitude of charge fluctuations in the simulation box.
Concerning the representation of the gauge fields, we identify the logarithmic encoding of the spin $S$ operators in the QLM as the most efficient one, and in particular an optimal setting is found when $2S+1 = 2^l$, with $l$ integer, which we denoted as \emph{perfectly representable} encoding.

Concerning the scaling of the Hamiltonian operator, the most expensive term is represented by the \emph{plaquette} term, that only exists in the multidimensional case.
This operator  requires a number of entangling gates which rapidly increases with the QLM truncation $S$.
Nevertheless, under the assumption discussed above, the number of Pauli terms to represent it, and the circuit depth to simulate it, both scale linearly with the volume.

However, for a precise assessment of the number of gates needed, information about the actual hardware connectivity is also required.
Nevertheless, even the time-evolution under the action of one Trotter step looks unfeasible for today's devices, as the resulting circuits feature $\mathcal{O}(10^3)$ CNOT gates even for relatively small systems. 

We stress that the large budget requirements that we report in this work are linked to the generality of the approach pursued, as in dimensions larger than one integrating-out gauge field degrees of freedom is not possible anymore (at variance with Ref.~\cite{Kokail2018Self-VerifyingModel}) and they remain therefore independent dynamical variables.
Moreover, to maintain the scalability of the approach, we choose to not rely on an exponentially expensive classical pre-processing to eliminate the non-physical sector of the Hilbert space before starting the simulation (at variance with Refs.~\cite{KlcoQuantum-ClassicalComputers,PhysRevA.100.012320}).

In this work, we mostly focused on the real-time evolution algorithm due to its importance for observing  non-equilibrium phenomena, which are hardly accessible in classical computations, and for calculating ground state properties using a QPE approach.
Our results are also relevant for the calculation of ground state properties using the VQE algorithm, which is best suited for current noisy quantum hardware and will be discussed in a following publication.

Finally, we also presented two test case simulations of a one and two-dimensional QED lattice model, showing the potential of this approach in describing interesting physics like string-breaking and confining on low dimensions. 
Even though a breakthrough in LGT will only be possible with the advent of fault tolerant quantum computers, these simple applications clearly indicate the potential of quantum computing in the domain of non-perturbative particle physics.
We believe that the resource count outlined in this work clearly shows that the simulation of LGT remains a challenging task, even when tackled using quantum computers. Further algorithmic developments and optimizations will be necessary to taper off such requirements toward the first real-time dynamics simulation of QED or QCD models, allowing physical predictions in the continuous limit.
Future research directions will include the generalization of this quantum computing framework and corresponding scaling laws to arbitrary $SU(N)$ gauge fields models.

%%%%%%%%%%%%%%%%%%%%%%%%%%%%%%%%%%%%%%%%%%%%%%%%%%%%%%%%%%%%%%%%%%%%%%%%%%%%%%%%%%%%%%%%%%%%%%%%%%%%%%%%%%%%%%%%%%%%%%%%%%%%%%%%%%%%%%%%%%%%%%%%%%%%%%%

%%%%%%%%%%%%%%%%%%%%%%%%%%%%%%%%%%%%%%%%%%%%%%%%%%%%%%%%%%%%%%%%%%%%%%%%%%%%%%%%%%%%%%%%%%%%%%%%%%%%%%%%%%%%%%%%%%%%%%%%%%%%%%%%%%%%%%%%%%%%%%%%%%%%%%%

%%%%%%%%%%%%%%%%%%%%%%%%%%%%%%%%%%%%%%%%%%%%%%%%%%%%%%%%%%%%%%%%%%%%%%%%%%%%%%%%%%%%%%%%%%%%%%%%%%%%%%%%%%%%%%%%%%%%%%%%%%%%%%%%%%%%%%%%%%%%%%%%%%%%%%%

\appendix

%\section{Fermion doubling problem}
%\label{app:doubling}
 
\section{Discretization of gauge fields}
\label{app:gauge}
For  small lattice spacing $a$ this term becomes
\begin{multline}
    \label{4:eq:plaquette_term}
    \hat  U_{(x, kj)} = \exp \big(-iqa \big( \hat  A_k(x) +  \\\hat A_j(x+k) - \hat  A_k(x_j) -  \hat A_j(x)\big) \big) \\
    = \exp \big(-iqa^2
    \big( \dfrac{ \hat A_j(x+k) -  \hat A_j(x)}{a} - \\\dfrac{ \hat A_k(x+j) - \hat  A_k(x)}{a}
    \big) \big)\\
    = 1 - iqa^2  \hat F_{jk}(x) + (iqa^2)^2  \hat F_{jk}(x)  \hat F^{jk}(x)/2! + \mathcal{O}({a^2}).
\end{multline}
   
We can thus reconstruct the magnetic field term in the Hamiltonian as 
\begin{equation}
    \label{4:eq:magnetic_field_contributionapp}
    \dfrac{1}{4 q^2 a^4} \sum\limits_{j<k} 2 - \left( \hat U_{x, kj} + \hat  U^\dagger_{x, kj} \right) = \dfrac{1}{4}  \hat F_{kj}(x) \hat  F^{kj}(x) 
    + \mathcal{O}({a^2})
\end{equation}
in the continuum limit.\\

\section{Example for mixed terms arising in non-perfect embedding}
\label{app:mixed}
In Section~\ref{ss:plaquette}, we reference Pauli terms with coefficient structure $(a+ia)$ in $\hat{U} \sim \hat{S}_x + i \hat{S}_y$. 
These arise due to shared terms in the $\hat{S}_x$ and $\hat{S}_y$ operators in a logarithmic spin-to-qubit-embedding, where $2S+1$ is not a power of two. Let us illustrate these terms by way of an example with $S=1$. 

In this case, the logarithmically embedded operators are 
\begin{align}
\hat{S}_x &= \dfrac{1}{4} (II - IZ - ZI + ZZ) \nonumber \\&+ \dfrac{1}{2 \sqrt{2}} (IX + XX + YY + ZX) \\ 
\hat{S}_y &= \dfrac{1}{4} (II - IZ - ZI + ZZ) \nonumber \\&+  \dfrac{1}{2 \sqrt{2}} (IY + YX - XY + ZY),
\end{align}
with the tensor product $\otimes$ suppressed for clarity. 
Upon summing $\hat{U} \sim \hat{S}_x + i \hat{S}_y$ there are four terms with coefficients $\pm \frac{1}{4}(1+i)$, namely
\begin{equation}
    II, IZ, ZI, ZZ.
\end{equation}
The terms with the structure $(a+ia)$ (in this case with $a \in {II,IZ,ZI,ZZ}$) lead to the purely imaginary contributions in $\hat{U}_\Box$.
These shared terms in $\hat{S}_x$ and $\hat{S}_y$ are a consequence of the %remaining computational basis state 
presence of `unused' states 
when the three eigenstates of a spin $S=3/2$ operator are %embedded 
mapped into the four computational basis states of two qubits. 
When the logarithmic embedding makes use of all states in the computational basis, no such terms with structure $(a+ia)$ arise.

\section{Circuit for exponential of Pauli operators}
\label{app:circuit}

\begin{figure}[htb]
    \centering
    \includegraphics{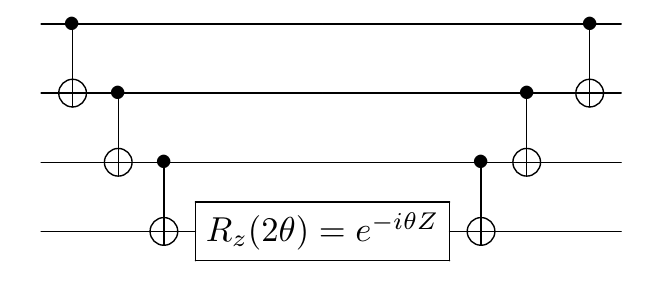}
    \caption{Circuit for exponentiating the Pauli string $\hat{H} = Z_1 Z_2 Z_3 Z_4 $ to $e^{-i\hat{H}\theta}$. 
    The first CNOT ladder computes the parity of qubits 1 to 4 and stores it in qubit 4. A phase shift 
    with the parameter $\theta$ is then applied to these qubits with a sign depending on the state of qubit 4, which encodes the parity.
    Finally, the second CNOT ladder uncomputes the first CNOT ladder.}
    \label{4:fig:pauli_z_exponential}
\end{figure}

To explain how Pauli strings can be exponentiated, let us start with a Pauli string composed of only $Z$ operators and then generalize to arbitrary Pauli strings.
Suppose we have the hermitian operator 
\begin{equation}
    \hat{H} = Z_1 Z_2 \dots Z_n .
\end{equation}
Then the operator $e^{-i\hat{H} \theta}$ adds a phase factor of $e^{-i \theta}$ if the parity in the computational basis of the $n$ qubits is even and a phase of $e^{i \theta}$ if it is odd. This operation can be implemented as a simple circuit by first computing the parity of the $n$ qubits in the computational basis, applying a phase shift of $e^{\pm i \theta}$ conditioned on the parity, and then uncomputing this phase shift again. The circuit in Fig.~ \ref{4:fig:pauli_z_exponential} performs exactly this operation for $n=4$. 

\begin{figure}[htb]
    \centering
    \includegraphics{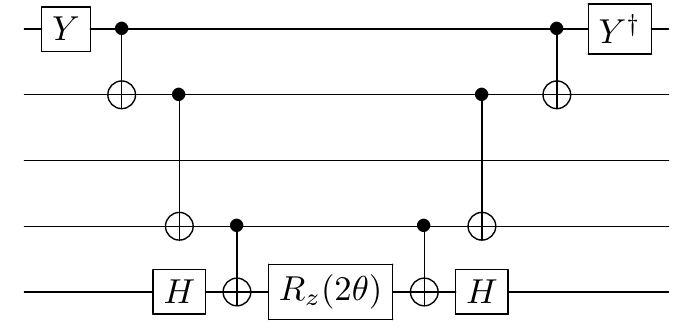}
    \caption{Circuit for exponentiating a Pauli string. This circuit exponentiates $\exp(i \theta Y_0 Z_1 Z_3 X_4)$.}
    \label{4:fig:pauli_exponential}
\end{figure}
The above procedure also allows us to exponentiate more complicated Pauli strings which contain $X$ and $Y$ operators, by simply performing a pre-rotation into the $X$ and $Y$ eigenbases on the respective qubits and then to evaluate the same circuit as in \ref{4:fig:pauli_z_exponential}. For example, the exponential $\exp(i \theta Y_0 Z_1 Z_3 X_4)$ can be calculated with the circuit in \ref{4:fig:pauli_exponential}.\\

\section{Resources for selected lattices}
\label{app:numbers}
In this Section we report precise qubits resources and number of Pauli operators counts to implement selected lattices and different QLM truncations.

\begin{table}[htb]
\centering
\ra{1.4}
\begin{tabular}{lrrrr}
\toprule
{} $S$ &  hopping term & $E$ op.  & $E^2$ term & plaquette term\\
\midrule
\hline
0.5   &                4  &   1  & 1 & 8\\
1.0   &               32 &  4  & 4 &  15616\\
1.5   &               12 &  2  & 2 & 648\\
2.0   &               80 &  8  & 8 & 772096\\
3.0   &               80 &  8  & 8 & 772096\\
3.5   &               32 &   3 & 4 & 32768\\
7.5   &               80 &  4  & 7 & 1280000 \\
15.5  &              192 &  5  & 11 & 42467328\\
31.5  &              448 &  6  & 16 &  1258815488\\
63.5  &             1024 &  7  & 22 & 34359738368\\
127.5 &             2304 &  8  & 29 & 9~ 10$^{11}$\\
255.5 &             5120 &   9  & 37 & 2~ 10$^{14}$\\
\hline
\bottomrule
\end{tabular}
\caption{Number of Pauli operators necessary to represent, respectively, the hopping term, the electic field $E$ operator (necessary for the Gauss Law), the electric field square $E^2$ term , and the plaquette term in the Hamiltonian, for several several QLM truncation $S$. Notice the drop in the resources required for the perfectly representable setting discussed in the main text.
The plaquette term dominates in the scaling with $S$.
To calculate the total number of Paulis is necessary to multiply these values for the total number of links and plaquettes present in the lattice.
}
\end{table}

\begin{table*}[htb]
\centering
\ra{1.4}
\begin{tabular}{llrrr}
\hline
\toprule
                &       &  nqubits\_total &  nqubits\_fermionic &  nqubits\_gauge \\
Lattice layout & S &                &                    &                \\
\midrule
\hline
(2, 3) & 0.5   &             19 &                 12 &              7 \\
                & 1.0   &             26 &                 12 &             14 \\
                & 1.5   &             26 &                 12 &             14 \\
                & 3.5   &             33 &                 12 &             21 \\
\hline
(4, 4) & 0.5   &             56 &                 32 &             24 \\
                & 1.0   &             80 &                 32 &             48 \\
                & 1.5   &             80 &                 32 &             48 \\

                & 3.5   &            104 &                 32 &             72 \\
                & 7.5   &            128 &                 32 &             96 \\
\hline
(10, 10) 
                & 1.0   &            560 &                200 &            360 \\
                & 1.5   &            560 &                200 &            360 \\
                & 3.5   &            740 &                200 &            540 \\
                & 7.5   &            920 &                200 &            720 \\
\hline
(100, 100)
                & 1.0   &          59600 &              20000 &          39600 \\
                & 1.5   &          59600 &              20000 &          39600 \\
                & 3.5   &          79400 &              20000 &          59400 \\
                & 7.5   &          99200 &              20000 &          79200 \\
                & 15.5  &         119000 &              20000 &          99000 \\
\hline
\hline
\bottomrule
\end{tabular}
\caption{Qubit register size needed for several 2D lattices and several QLM truncation $S$}
\end{table*}

\begin{table*}[htb]
\centering
\ra{1.4}
\begin{tabular}{llrrr}
\hline
\toprule
                &       &  nqubits\_total &  nqubits\_fermionic &  nqubits\_gauge \\
Lattice layout & S &                &                    &                \\
\midrule
\hline
(2, 2, 2) & 0.5   &             44 &                 32 &             12 \\
                & 1.0   &             56 &                 32 &             24 \\
                & 1.5   &             56 &                 32 &             24 \\
                & 3.5   &             68 &                 32 &             36 \\
\hline
(4, 4, 4)                 & 1.0   &            544 &                256 &            288 \\
                & 1.5   &            544 &                256 &            288 \\
                & 3.5   &            688 &                256 &            432 \\
                & 7.5   &            832 &                256 &            576 \\
                & 15.5  &            976 &                256 &            720 \\
\hline
(10, 10, 10)               & 1.0   &           9400 &               4000 &           5400 \\
                & 1.5   &           9400 &               4000 &           5400 \\
                & 7.5   &          14800 &               4000 &          10800 \\
                & 15.5  &          17500 &               4000 &          13500 \\
                & 31.5  &          20200 &               4000 &          16200 \\
\hline
(100, 100, 100) 
                & 1.0   &        9940000 &            4000000 &        5940000 \\
                & 1.5   &        9940000 &            4000000 &        5940000 \\
                & 3.5   &       12910000 &            4000000 &        8910000 \\
                & 7.5   &       15880000 &            4000000 &       11880000 \\
                & 15.5  &       18850000 &            4000000 &       14850000 \\
                & 127.5 &       27760000 &            4000000 &       23760000 \\
                & 255.5 &       30730000 &            4000000 &       26730000 \\
\hline
\bottomrule
\end{tabular}
\caption{Qubit register size needed for several 3D lattices and several QLM truncation $S$}
\end{table*}

%\bibliography{main}
\bibliography{references,bib_qc}
\end{document}